\documentclass[aps,pre,twocolumn,showpacs,superscriptaddress,groupedaddress]{revtex4-2}
\usepackage{amsmath}
\usepackage{amsfonts}
\usepackage{amssymb}
\usepackage{graphicx}
\usepackage{dcolumn}
\usepackage{sidecap}
\usepackage[dvipsnames]{xcolor}
\usepackage[colorlinks=true,allcolors=blue]{hyperref}
\usepackage{hhline}
\usepackage{mathtools}
\usepackage{multirow}
\usepackage{verbatim}
\usepackage{minitoc}
\usepackage{rotating}
\usepackage{setspace}
\usepackage{epsfig}
\usepackage{epstopdf}
\usepackage{dsfont}

\usepackage[ruled]{algorithm2e}

\usepackage{dcolumn}
\usepackage{bm}
\usepackage{amsfonts}
\usepackage{bbm}
\usepackage{booktabs}
\usepackage{lipsum}
\usepackage[normalem]{ulem}

\usepackage{soul}
\usepackage{bm}

\usepackage{lettrine}

\makeatletter
\def\@eqnnum{{\normalsize \normalcolor (\theequation)}}
 \makeatother

\begin{document}

\title{Percolation in the Stochastic Block Model}

\author{Luca~Pieleanu}\thanks{\href{mailto:l.pieleanu@northeastern.edu}{l.pieleanu@northeastern.edu}}
\affiliation{Sharon High School, Sharon, Massachusetts, USA 02067}
\affiliation{Network Science Institute, Northeastern University, Boston, Massachusetts, USA 02115}

\author{Moritz~Laber}
\affiliation{Network Science Institute, Northeastern University, Boston, Massachusetts, USA 02115}
\affiliation{Complexity Science Hub, Vienna 1030, Austria}

\author{Narayan~G.~Sabhahit}
\affiliation{Network Science Institute, Northeastern University, Boston, Massachusetts, USA 02115}

\author{Dmitri~Krioukov}\thanks{\href{mailto:dima@northeastern.edu}{dima@northeastern.edu}}
\affiliation{Network Science Institute, Northeastern University, Boston, Massachusetts, USA 02115}
\affiliation{Department of Physics, Northeastern University, Boston, Massachusetts, USA 02115}
\affiliation{Department of Mathematics, Northeastern University, Boston, Massachusetts, USA 02115}
\affiliation{Department of Electrical \& Computer Engineering, Northeastern University, Boston, Massachusetts, USA 02115}

\date{\today}

\begin{abstract}
The stochastic block model is a paradigmatic model of networks with community structure. Yet percolation in the model has been studied primarily in cases with a fixed block structure, even though in real networks, the community structure may evolve as the network grows. Here we study percolation in sequences of stochastic block models in which the numbers and sizes of communities, as well as the intra- and intercommunity connection probabilities may all change with the network size. We analyze five such sequences using two methods: linearized self-consistent equations for the locally tree-like models and a branching process at the community scale for the models with nonvanishing clustering. We find that the critical average degree is not generally equal to $1$, even in locally tree-like sequences, because the transition depends on how connections are distributed across the evolving community structure. We also show that the community-scale branching process accurately predicts the transition when intercommunity connections are sufficiently sparse, even in the presence of nonvanishing clustering, while a geometric stochastic block model sequence demonstrates the limitations of this method when correlations between intercommunity connections cannot be neglected. These results extend percolation studies in the stochastic block model to more realistic scenarios with evolving community structure and may provide new methods to derive the upper and lower bounds for the percolation threshold in geometric long-range percolation.
\end{abstract}

\maketitle

\section{Introduction}\label{sec:introduction}
Percolation studies how large-scale connectivity emerges from local connections in a network. As connectivity increases, a network can pass from a collection of small connected components to a regime containing a giant connected component (GCC), which contains a nonvanishing fraction of all vertices. Because a GCC allows a process to spread across a large part of a network, its emergence is central to models of epidemic spreading~\cite{newman2002_SpreadEpidemicDisease} and to studies of network robustness under node or edge failures~\cite{callaway2000_NetworkRobustnessFragility, artime2024_RobustnessResilienceComplex, jiang2025_RobustnessSmallNetworks}. Percolation theory hence provides a common framework for studying these phenomena~\cite{li2021_PercolationComplexNetworks}. In the classical Erdős–Rényi random graph model, the GCC emerges when the average degree reaches one~\cite{erdos1960_evolutionrandomgraphs}. This condition provides a useful reference point, but it is not universal, as degree heterogeneity can shift the transition away from an average degree of one~\cite{molloy1995_criticalpointrandom, newman2000_Randomgraphsarbitrary, cirigliano2024_ScalingUniversalityPercolation}, while type-dependent mixing can make the threshold depend on the full pattern of connections between different types of vertices~\cite{newman2003_MixingPatternsNetworks,allard2009_HeterogeneousBondPercolation}. At a larger scale, the arrangement of communities and other network structures can likewise alter connectivity and robustness~\cite{morelbalbi2022_LargeScaleStructure, morel-balbi2020_NullModelsMultioptimized, peixoto2012_EvolutionRobustNetwork, allard2019_PercolationEffectiveStructure}. These results show why the average degree alone is generally insufficient to locate the transition.

Clustering introduces a related but distinct difficulty. Many standard percolation calculations assume that local neighborhoods become tree-like as the network grows~\cite{karrer2014_Percolationsparsenetworks, newman2023_MessagePassingMethods}, but the presence of short loops in a network can violate the independence underlying these calculations and can therefore change their predictions~\cite{cantwell2019_MessagePassingLoops, radicchi2016_BeyondLocallyTreelike}. The resulting effect on the percolation threshold has been analyzed in several clustered random graph models~\cite{miller2009_Percolationepidemicsrandom, gleeson2009_AnalyticalResultsBond, gleeson2009_Howclusteringaffects}. Exact treatments have also been developed for solvable ensembles constructed from triangles or more general small subgraphs~\cite{newman2009_RandomGraphsClustering, karrer2010_Randomgraphscontaining, allard2015_GeneralExactApproach}.

The stochastic block model (SBM) provides a natural framework for studying how group structure affects network connectivity because its connection probabilities may vary between different parts of a network while the model remains analytically tractable. The SBM was introduced as a general description of block-structured networks~\cite{holland1983_StochasticBlockmodels} and can approximate broader classes of network structure~\cite{olhede2014_NetworkHistogramsUniversality}. Sparse SBMs, including planted-partition models, have become a canonical setting for community detection and its statistical and computational transitions~\cite{karrer2011_StochasticBlockmodelsCommunity, abbe2018_CommunityDetectionStochastic, decelle2011_AsymptoticAnalysisSBM, bhamidi2026_StochasticBlockModel}. From the perspective of connectivity, closely related finite-type random graphs have a well-developed theory of phase transitions and component formation~\cite{bollobas2005_phasetransitioninhomogeneous, soderberg2002_Generalformalisminhomogeneous, kang2015_phasetransitionmultitype}. Most of these asymptotic treatments keep the number of blocks and their limiting relative sizes fixed as the network grows. They therefore do not describe what happens when the block structure itself changes with network size.

Here, we address this question by studying \emph{sequences} of SBMs (SBMSs) in which the number of communities, their sizes, and both the intra- and intercommunity connection probabilities may all vary with $n$. We consider five sequences that illustrate how different scalings of these quantities affect percolation. The first is precisely the classical sparse fixed-block regime mentioned above. Its percolation and component structure have been analyzed extensively~\cite{bollobas2005_phasetransitioninhomogeneous, soderberg2002_Generalformalisminhomogeneous, kang2015_phasetransitionmultitype, bujok2014_PercolationClassicalBlockmodel, schawe2020_LargeDeviationsConnected}, most recently in~\cite{franchi2026_ComponentStructurePercolation}. In the second sequence, the number of blocks grows with $n$, while in the third, the communities correspond to the layers of a perfect $a$-ary tree. While these first three sequences have locally tree-like behavior, the remaining two sequences do not, retaining nonvanishing clustering as the network grows. The fourth consists of a growing number of fixed-size dense planted Erdős–Rényi communities connected by sparse edges. Each graph in the sequence has the planted-partition form widely studied in the SBM literature~\cite{karrer2011_StochasticBlockmodelsCommunity, abbe2018_CommunityDetectionStochastic, decelle2011_AsymptoticAnalysisSBM}, but here the number of communities grows with $n$ while their sizes remain fixed. Its percolation behavior is therefore more closely related to the classical blockmodel~\cite{bujok2014_PercolationClassicalBlockmodel} and to random graph models built from finite households or communities~\cite{ball2008_Thresholdbehaviourfinal, vanderhofstad2016_HierarchicalConfigurationModel, vanderhofstad2022_PhaseTransitionRandom, stegehuis2016_PowerLawRelationsCommunities}. The fifth is a geometric SBMS obtained by averaging a soft random geometric connection kernel over blocks. This relates it to continuum percolation, soft random geometric graph theory, and latent-space network models~\cite{penrose1991_continuumpercolationmodel, penrose2022_Giantcomponentsoft, penrose2016_ConnectivitySoftRandom, hutchcroft2025_CriticalLongrangePercolation1, hutchcroft2025_CriticalLongrangePercolation2, hutchcroft2025_CriticalLongrangePercolation3, hutchcroft2021_Powerlawboundscritical, hoff2002_LatentSpaceApproaches}. Together, these five sequences move from the standard regime of a fixed number of blocks to settings in which the community structure evolves along with the network.

The distinction between the locally tree-like and clustered sequences determines our analytical approach. For the first three sequences, we locate the transition by linearizing self-consistency equations for GCC membership. This produces a threshold determined by the largest eigenvalue of the matrix of expected connections between communities. For the two clustered sequences, we first determine the connected structure within each community and then study how those components connect across communities. We compare both analytical predictions with numerical simulations.

We find that the critical average degree is not generally equal to one, even for SBM sequences that become tree-like as $n$ grows. Whether the Erdős–Rényi condition is recovered depends on how connections are distributed across the communities and on how that structure scales with the size of the network. We also find that nonzero clustering does not by itself rule out a simple analytical description of the transition: when the connections between communities remain sufficiently sparse, the dense structure inside each community can instead be treated as part of a larger community-scale branching process. However, the geometric SBM shows the limitation of this picture when intercommunity connections are not sparse enough. More generally, these examples show that the percolation threshold of an SBM sequence is determined by its asymptotic community structure rather than by average degree or clustering alone.

The rest of the paper is organized as follows. In Sec.~\ref{sec:sbm-properties}, we define the general SBM sequence and derive expressions for its average degree and average local clustering coefficient. In Sec.~\ref{sec:methods}, we introduce the linearized self-consistency and community-scale branching-process methods, along with the numerical protocol used to locate the percolation threshold in simulations. Sections \ref{sec:zero-clustering-sbms} and \ref{sec:nonzero-clustering-sbms} apply these methods to the three locally tree-like SBM sequences and two clustered SBM sequences, respectively, and Sec.~\ref{sec:conclusion} summarizes the results and discusses the limitations of the two approaches.

\section{Stochastic Block Model and Its Properties}\label{sec:sbm-properties}

\begin{figure}
    \centering
    \includegraphics[width=\linewidth, trim={0cm 6cm 0cm 6cm}, clip]{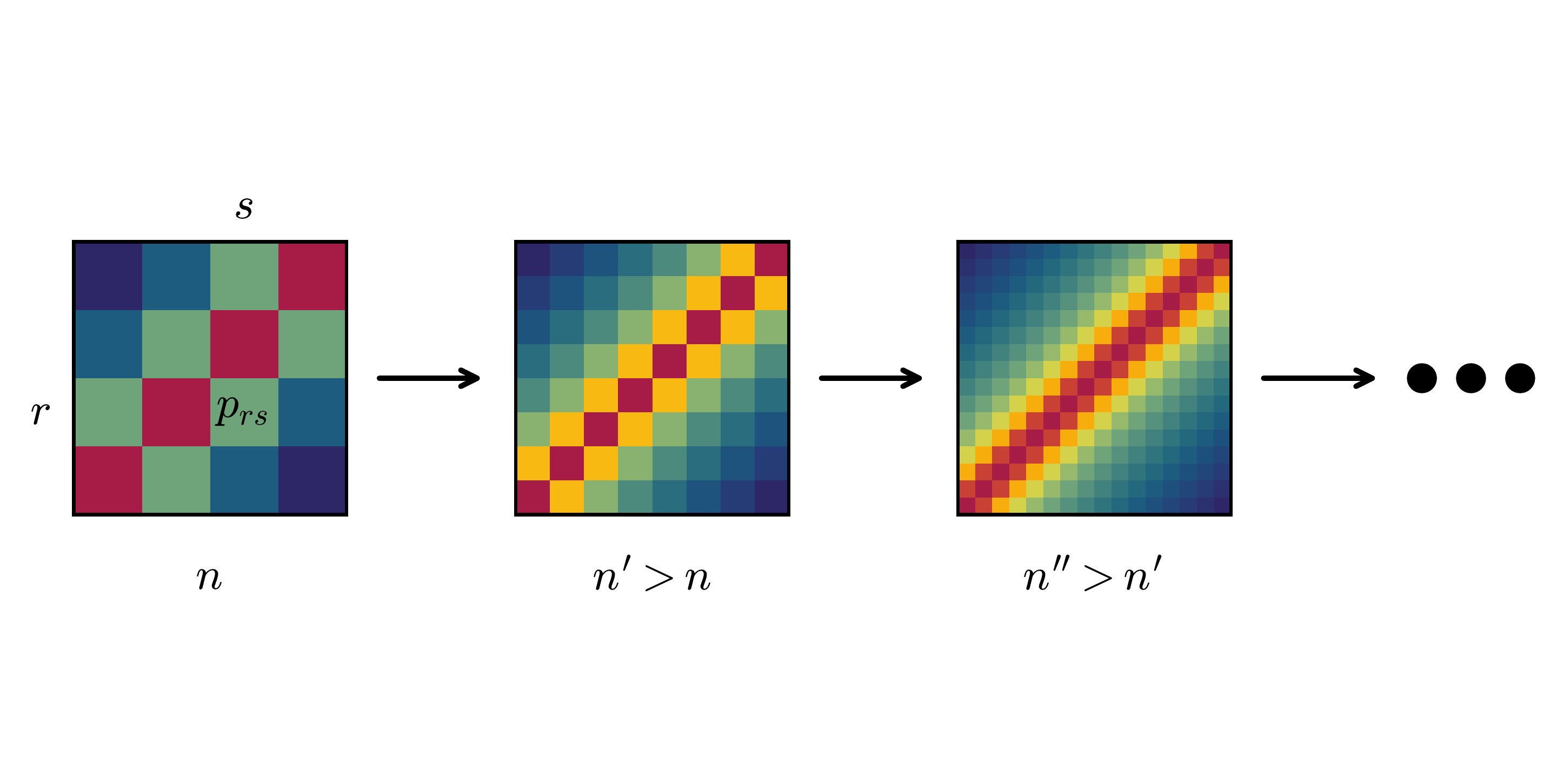}
    \caption{\textbf{Connection-probability maps for stochastic block model sequences (SBMSs).} Each grid represents the connection-probability matrix of an SBM in the sequence at a different graph size $n$. Rows and columns correspond to communities, and the color of the cell in row $r$ and column $s$ gives the connection probability $p_{rs}(n)$ between vertices in communities $r$ and $s$. The width of row or column $r$ is proportional to the relative size $n_r(n)/n$, where $n_r(n)$ is the size of community $r$, while the total number of rows or columns equals the number of communities $q(n)$. Different SBM sequences are characterized by different scalings of $q(n)$, $n_r(n)$, and $p_{rs}(n)$ with $n$, which determine their asymptotic structure and percolation behavior.}
    \label{fig:graphon-sequence}
\end{figure}

A stochastic block model is a random graph model in which vertices are divided into communities, and edge probabilities depend only on the communities of the two endpoints. For a fixed total number of vertices $n$, we take $q(n)$ communities $1, 2, \ldots, q(n)$, with sizes $n_1(n), n_2(n), \ldots, n_{q(n)}(n)$, respectively, where the community sizes sum to $n$. Then, for any two distinct vertices, one in community $r$ and one in community $s$, the edge between them is present independently with probability $p_{rs}(n)$. The case $r=s$ describes edges within the same community, while the case $r \neq s$ describes edges between distinct communities. Throughout this paper, we visualize SBM sequences using connection-probability maps, with a sample sequence shown in Fig.~\ref{fig:graphon-sequence}.

As stated in the introduction, in this paper we study stochastic block model \textit{sequences} rather than a single stochastic block model. As $n \to \infty$, the number of communities $q(n)$, the community sizes $n_r(n)$, and the connection probabilities $p_{rs}(n)$ may all vary with $n$. This is important because the asymptotic scaling of these quantities can substantially change the local structure of the graph, thus requiring different methods to identify the corresponding percolation threshold.

Before analyzing the five specific SBM sequences, we first derive two general structural quantities for the stochastic block model: the average degree $\bar{k}$ and the average local clustering coefficient $\bar{c}$. The average degree will be used to compare the percolation threshold with the condition $\bar{k}=1$, while the average local clustering coefficient will be used as a diagnostic of local tree-likeness at the level of triangles. For the first three sequences, we establish local tree-likeness directly and use $\bar{c} \to 0$ to verify the disappearance of triangles, while for the last two, a nonvanishing limit rules out a locally tree-like approximation.  

For a vertex in community $r$, the expected number of neighbors in community $s \neq r$ is $p_{rs}n_s$, while the expected number of neighbors in its own community is $p_{rr}(n_r-1)$, since the vertex cannot connect to itself. Therefore, the expected degree of a vertex in community $r$ is $p_{rr}(n_r-1)+\sum_{s \neq r} p_{rs}n_s$. Averaging this over all communities (weighted by the fraction of vertices in each community) gives
\begin{equation}\label{eq:general-sbm-average-degree}
    \bar{k} = \frac{1}{n} \sum_{r=1}^q n_r \left(p_{rr}(n_r-1)+\sum_{s \neq r} p_{rs}n_s\right).
\end{equation}
This expression gives the expected average degree of the general stochastic block model. In the SBM sequences considered below, we will evaluate this formula under the corresponding scalings of $q, n_r$, and $p_{rs}$, and compare the result with the percolation threshold predicted analytically.

Next, we compute the average local clustering coefficient $\bar{c}$. It is obtained by averaging the local clustering coefficient---the fraction of pairs of the neighbors of a vertex that are connected by an edge---over the vertices of degree at least $2$. Suppose that a vertex in community $r$ has $x_s$ neighbors in community $s$. Then, its total degree is $\sum_{s=1}^q x_s$. Conditioned on these neighbor counts, two neighbors both in community $s$ are connected with probability $p_{ss}$, while two neighbors in distinct communities $s$ and $t$ are connected with probability $p_{st}$. Therefore, conditioned on the values $x_1, \ldots, x_q$, the expected local clustering coefficient of the vertex is
\begin{equation}\label{eq:general-sbm-local-clustering-conditioned}
    \bar{c}(x_1, \ldots, x_q) = \frac{\sum_s p_{ss} \tbinom{x_s}{2}+\sum_{s<t} p_{st} x_s x_t}{\binom{\sum_{s} x_s}{2}},
\end{equation}
whenever $\sum_{s=1}^q x_s \ge 2$. To obtain $\bar{c}$, it remains to average this expression over the possible neighbor counts and over the choice of the root community, while conditioning on the root degree being at least $2$ for the local clustering coefficient to exist:
\begin{equation}\label{eq:general-sbm-local-clustering-unconditioned}
    \bar{c} = \mathbb{E}\left[ \frac{\sum_s p_{ss} \tbinom{x_s}{2} + \sum_{s <t} p_{st}x_s x_t}{\tbinom{\sum_s x_s}{2}} \ \middle| \ \sum_s x_s \ge 2\right],
\end{equation}
where, for a root vertex in community $r$, the neighbor counts satisfy $x_r \sim \text{Bin}(n_r-1, p_{rr})$ and $x_s \sim \text{Bin}(n_s, p_{rs})$, where $\text{Bin}$ stands for the binomial distribution. This expression gives the average local clustering coefficient of the general stochastic block model. In the SBM sequences considered below, we will evaluate this formula under the corresponding scalings of $q, n_r$, and $p_{rs}$ to obtain their limiting average local clustering coefficients.

Illustrating our theoretical derivations through simulations requires sampling graphs from the SBM. To sample these graphs efficiently, we adapt the Miller--Hagberg (MH) algorithm~\cite{miller2011_efficientgenerationnetworks} for the Chung--Lu model~\cite{chung2002_averagedistances} to the SBM.
Appendix~\ref{app:algorithm} provides pseudocode for the resulting algorithm and verifies that it samples from the intended network ensemble.

\section{Methods}\label{sec:methods}

We use two analytical methods to determine the percolation thresholds of the SBM sequences considered below, and we compare their predictions with numerical simulations. Here, the percolation threshold refers to the boundary in parameter space separating the regime in which the largest connected component contains a vanishing fraction of the vertices from the regime in which it contains a positive fraction with high probability as $n \to \infty$. This boundary may be expressed as an implicit relation among the model parameters or, when all parameters except a scalar control parameter $\nu$ are fixed, as a critical value $\nu_c$. The latter one-parameter representation is used in our numerical simulations, as discussed in Sec.~\ref{sec:numerics}.

The first method applies when local neighborhoods are asymptotically tree-like. In this setting, the emergence of the GCC can be studied through self-consistent equations for the probabilities that vertices belong to the GCC. Linearizing these equations around the trivial no-GCC solution to test its stability yields the standard spectral transition criterion for sparse, locally tree-like models, including configuration models, multitype random graphs, and inhomogeneous random graphs~\cite{newman2000_Randomgraphsarbitrary, karrer2014_Percolationsparsenetworks, newman2023_MessagePassingMethods, allard2009_HeterogeneousBondPercolation, bollobas2005_phasetransitioninhomogeneous, soderberg2002_Generalformalisminhomogeneous}.

The second method applies when short cycles persist in the neighborhoods of vertices, making the independence assumption underlying the first method inaccurate. When intercommunity edges are sparse, however, connectivity can instead be treated at two different scales: we first determine the connected structure within each finite community and then study how the resulting local components connect across communities. The same separation underlies threshold and giant-component analyses for household and community-structured random graphs~\cite{ball2008_Thresholdbehaviourfinal, vanderhofstad2016_HierarchicalConfigurationModel, stegehuis2016_PowerLawRelationsCommunities, vanderhofstad2022_PhaseTransitionRandom}.

We discuss these two methods, followed by the numerical protocol, in more detail below.

\subsection{Linearized self-consistency method}\label{sec:linearized-self-consistency}

\begin{figure}
    \centering
    \includegraphics[width=\linewidth, trim={8.25cm 2cm 8.25cm 0cm}, clip]{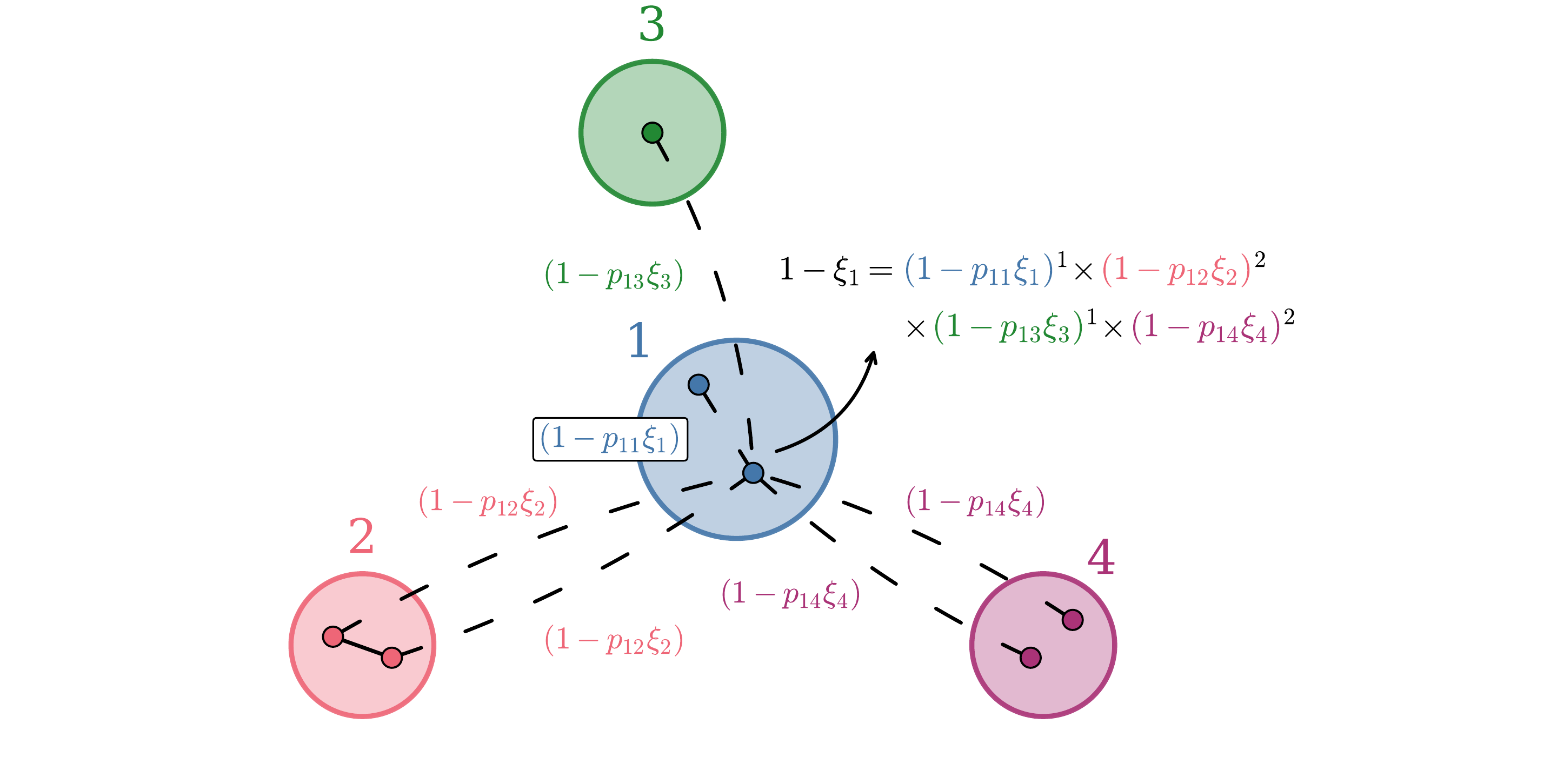}
    \caption{\textbf{Self-consistency equation.} A root vertex in community $1$ reaches the giant connected component (GCC) through a potential neighbor in community $s$ with probability $p_{1s}\xi_s$ for all $s \in \{1, 2, 3, 4\}$, where $\xi_s$ is the probability that a vertex in community $s$ belongs to the GCC. Consequently, $1-p_{1s}\xi_s$ is the probability that this neighbor does not connect the root vertex to the GCC. Assuming these events are asymptotically independent, which is valid when local neighborhoods are asymptotically tree-like, multiplying over all potential neighbors yields the probability $1-\xi_1$ that the root vertex does not belong to the GCC, giving the self-consistency equation \eqref{eq:fixed-point-xi}.}
    \label{fig:method1}
\end{figure}

Consider an SBM sequence whose finite-radius neighborhoods around a uniformly chosen vertex are asymptotically tree-like. Let $\xi_r$ be the probability that a vertex in community $r$ belongs to the GCC. Then, $1-\xi_r$ is the probability that such a vertex \textit{does not} belong to the GCC. Averaging over the community of the vertex, the corresponding overall GCC membership probability is
\begin{equation}
    \xi = \sum_{r=1}^{q}\frac{n_r}{n}\xi_r.
\end{equation}
Equivalently, $\xi$ is the predicted fraction of vertices in the GCC.

Fix a vertex in community $r$. For a vertex in community $s$, the fixed vertex reaches the GCC through that vertex if and only if the edge between them is present and the vertex in community $s$ belongs to the GCC. This occurs with probability $p_{rs}\xi_s$. Hence, the probability that the fixed vertex of community $r$ does not reach the GCC through this particular vertex of community $s$ is $1-p_{rs}\xi_s$. Since there are $n_r-1$ other vertices in community $r$ and $n_s$ vertices in each community $s \neq r$, the self-consistent equation for $\xi_r$ is
\begin{equation}\label{eq:fixed-point-xi}
    \xi_r = 1 - (1-p_{rr}\xi_r)^{n_r-1}\prod_{s \neq r} (1-p_{rs}\xi_s)^{n_s}.
\end{equation}
The construction of this equation is illustrated in Fig.~\ref{fig:method1}.

The solution $\xi_1 = \cdots = \xi_q=0$ corresponds to the absence of a GCC. To determine when a nonzero solution first appears, we examine when this fixed point loses stability.

To find this point, we linearize the self-consistent equation near $\xi_1 = \cdots = \xi_q=0$, assuming that each $\xi_r$ is small. Expanding Eq.~\eqref{eq:fixed-point-xi} to first order in the variables $\xi_r$ yields
\begin{equation}\label{eq:linearized-xi}
    \xi_r \approx p_{rr}(n_r-1)\xi_r + \sum_{s \neq r} p_{rs}n_s \xi_s.
\end{equation}
Let $t$ be the $q \times q$ matrix with diagonal terms $t_{rr}=(n_r-1)p_{rr}$, corresponding to expected connections from a vertex in community $r$ to other vertices in the same community, and off-diagonal terms $t_{rs}=n_sp_{rs}$ for $s \neq r$, corresponding to expected connections from a vertex in community $r$ to vertices in community $s$. Also, let $\boldsymbol{\xi}$ denote the column vector with entries $\xi_1, \ldots, \xi_q$. Then, the linearized system can be written as
\begin{equation}\label{eq:self-consistent-equation}
    \boldsymbol{\xi} \approx t\boldsymbol{\xi}.
\end{equation}
Near the trivial solution, each iteration of the self-consistency equations approximately replaces $\boldsymbol{\xi}$ by $t\boldsymbol{\xi}.$ Decomposing $\boldsymbol{\xi}$ into eigenvector components, each component is multiplied by its corresponding eigenvalue. Since $t$ has nonnegative entries, its largest eigenvalue $\lambda_{\mathrm{max}}(t)$ also has the largest absolute value. Thus, all components decay when $\lambda_{\mathrm{max}}(t)<1,$ while a component along the leading eigenvector grows when $\lambda_{\mathrm{max}}(t)>1.$ The stability threshold is therefore
\begin{equation}\label{eq:fixed-point-threshold}
    \lambda_{\max}(t)=1.
\end{equation}
This spectral condition is the prediction of the linearized self-consistency method for the percolation threshold. When $q(n)$ grows, however, checking when the trivial fixed point $\boldsymbol{\xi} = \mathbf{0}$ loses stability is not sufficient to determine the percolation threshold. In particular, a nonzero $\xi_r$ contributes only $(n_r/n)\xi_r$ to $\xi$, so nonzero coordinates confined to communities whose total relative size tends to zero may still yield $\xi \to 0$. Whenever we apply this method to a sequence with a growing number of communities, i.e., in Sec.~\ref{sec:growing-block-sbm} and Sec.~\ref{sec:tree-layers-sbm}, we therefore verify analytically that $\xi$ remains bounded away from zero above the predicted threshold. Together with the asymptotic independence assumption underlying Eq.~\eqref{eq:fixed-point-xi}, which is valid for the locally tree-like sequences considered below, this ensures that the loss of stability corresponds to the emergence of a GCC.

\subsection{Community-scale branching process method}\label{sec:community-branching-process}

When short cycles persist in the neighborhoods of vertices in the thermodynamic limit ($n \to \infty$), the linearized self-consistent equations above can give the wrong transition point because different paths from a fixed vertex can pass through the same community, so multiplying the factors in the fixed-point equation no longer captures the correct branching structure. In this setting, we first analyze the connected component formed inside one community, and then count the intercommunity edges leaving it, in order to determine the effective branching factor between communities.

Choose a vertex uniformly at random and ignore all intercommunity edges. Let $T$ be the number of vertices reached inside the starting community, i.e., the size of the connected component containing the starting vertex when only intracommunity edges are kept. Also, let $\bar{k}_{\mathrm{out}}$ be the expected number of intercommunity edges leaving a single vertex. If intercommunity edges are sparse, then, conditional on $T$, the number of outgoing edges from this connected component is approximately Poisson distributed with mean $\bar{k}_{\mathrm{out}}T$.

\begin{figure}
    \centering
    \includegraphics[width=\linewidth, trim={5cm 3cm 4.75cm 2.5cm}, clip]{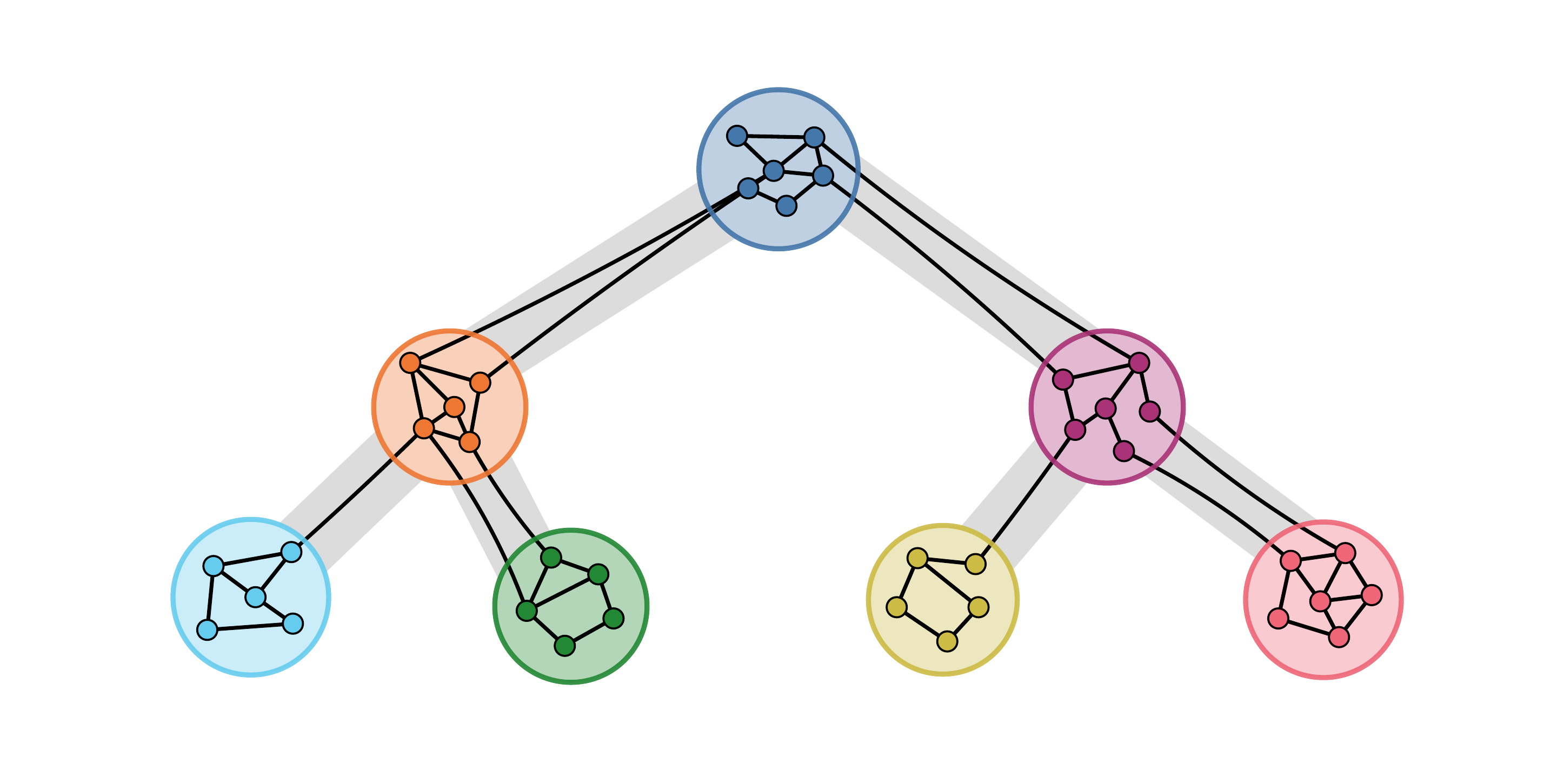}
    \caption{\textbf{Community-scale branching process.} Each colored disk represents a community after ignoring intercommunity edges, with the internal graph containing the connected component of a randomly chosen vertex. If $T$ denotes the size of this component and $\bar{k}_{\mathrm{out}}$ is the expected number of intercommunity edges leaving a single vertex, then a component generates, on average, $T  \bar{k}_{\mathrm{out}}$ outgoing intercommunity edges. Treating these sparse intercommunity connections as a branching process yields the condition $\mathbb{E}[T]\bar{k}_{\mathrm{out}} > 1$ in order for the branching process to survive with positive probability and hence the GCC to form, which remains valid even though the local neighborhoods inside communities are not tree-like.}
    \label{fig:method2}
\end{figure}

Let $\mu$ be the probability that the intracommunity component containing this uniformly chosen vertex does not lead to the GCC. Thus, $1-\mu$ is the predicted fraction of vertices in the GCC. Each outgoing edge leads to the GCC with probability $1-\mu$, so, conditional on $T,$ the number of such edges is approximately Poisson distributed with mean $\bar{k}_{\mathrm{out}}(1-\mu)T.$ The probability that none leads to the GCC is therefore $e^{-\bar{k}_{\mathrm{out}}(1-\mu)T}$. Then, averaging over $T$ gives
\begin{equation}\label{eq:fixed-point-mu}
    \mu = \mathbb{E}\left[ e^{-\bar{k}_{\mathrm{out}}(1-\mu)T}\right].
\end{equation}
The trivial solution $\mu=1$ corresponds to the absence of a GCC. Near the threshold, $\mu = 1 - \epsilon$, where $\epsilon \ll 1$, leading to
\begin{equation}\label{eq:linearized-mu}
    1 - \epsilon = \mathbb{E}\left[e^{-\bar{k}_{\mathrm{out}}\epsilon T}\right] = 1 - \epsilon\bar{k}_{\mathrm{out}} \mathbb{E}[T] + o(\epsilon).
\end{equation}
Therefore, the threshold condition is
\begin{equation}\label{eq:community-branching-threshold}
    \mathbb{E}[T]\bar{k}_{\mathrm{out}} = 1,
\end{equation}
as illustrated in Fig.~\ref{fig:method2}. This condition says that the GCC emerges when one connected component inside a community produces, on average, one outgoing intercommunity connection leading to another such component. The factor $\mathbb{E}[T]$ captures the connected structure inside communities, while $\bar{k}_{\mathrm{out}}$ captures the sparse connections between communities.

In the clustered SBM sequences considered below, these two quantities are computed from the specific scaling of the SBMS.

\subsection{Numerical methods}\label{sec:numerics}

We also confirm our analytical results in simulations. In simulations, we determine the percolation threshold by varying a scalar control parameter $\nu$ that influences the connection probability $p_{rs}(n)$ at a given number of nodes $n$. The choice of control parameter depends on the type of SBM sequence studied, e.g., $p_{ss}=\nu/n$ for all $1\leq s\leq q(n)$ would be a valid choice. 

At each value of $n$ in the SBM sequence, we vary the control parameter $\nu$ in a fixed interval $[\nu_{\mathrm{min}},\nu_{\max}]$ and sample $n_G$ graphs at each value of $\nu$.

For each graph, we use the tree data structure underlying the Newman--Ziff algorithm~\cite{newman2000_efficientmontecarlo, newman2001_fastmontecarlo} to efficiently determine the size of all connected components $C$ of the graph $G$. We denote the set of $G$'s connected components as $\mathcal{C}(G)$, and the largest component among them $C_{\max}(G)$. If the maximum is not unique, we break ties arbitrarily.

For the set of graphs $\mathcal{G}(\nu)=\{G_i\}_{i=1}^{n_G}$ each with $n$ nodes and sampled at control parameter value $\nu$, we record the average relative size of the largest connected component,
\begin{equation}\label{eq:exp-xi-def}
    \xi_\mathrm{exp}(\nu)
    =
    \frac{1}{n_G}
    \sum_{i=1}^{n_G}
    \frac{\left| C_{\max}(G) \right|}{n},
\end{equation}
and the average susceptibility, i.e., the expected size of a nonlargest component containing a uniformly chosen vertex,
\begin{equation}\label{eq:exp-chi-def}
    \chi_\mathrm{exp}(\nu)
    =
    \frac{1}{n_G}
    \sum_{i=1}^{n_G}
    \sum_{C \in \mathcal{C}(G)\setminus C_{\max(G)}}
    \frac{|C|^2}{n}
    .
\end{equation}
We then take the value $\nu_c$ of the control parameter for which $\chi_\mathrm{exp}$ reaches its maximum as a finite-size estimate of the asymptotic critical value, i.e.,
\begin{equation}
    \nu_c = \arg \max\{\chi_{\mathrm{exp}}(\nu) : \nu \in [\nu_{\mathrm{min}},\nu_{\max}]\}.
\end{equation}

In all numerical simulations, we use $n_G=100$ sampled graphs. For visualization, we use the one-to-one correspondence between the control parameter $\nu$ and the average degree $\bar{k}$, with all other parameters held fixed.

For the numerical examples whose thresholds are characterized by the condition $\lambda_{\max}=1$, the matrices $t$ possess sufficient structure to admit closed-form expressions for their largest eigenvalue. Therefore, the reported values of $\lambda_{\max}$ are obtained by evaluating the corresponding closed-form expressions for $t$ at the critical value of the control parameter $\nu$. These expressions are derived in the corresponding sections.

\section{Locally tree-like SBM Sequences}\label{sec:zero-clustering-sbms}

We first consider three SBM sequences whose finite-radius neighborhoods around a uniformly chosen vertex are asymptotically tree-like. We establish this property directly for each sequence below, allowing the linearized self-consistency method from Sec.~\ref{sec:linearized-self-consistency} to be applied. Despite sharing this locally tree-like behavior, the three sequences differ in how their communities scale with $n$ and how edges are distributed among them.

The first SBMS is the classical sparse SBM sequence with a constant number of blocks, where the number of communities is fixed, the community sizes grow linearly with $n$, and all connection probabilities decay as $1/n$. This family includes several sparse random graph models as special cases. With one block, it reduces to the Erdős–Rényi graph $G(n, c/n)$. With two blocks and no intrablock edges, it becomes the bipartite Erdős–Rényi model, for which the percolation threshold and the behavior of the largest component near the transition have been studied~\cite{johansson2012_giantcomponentrandom, kang2015_phasetransitionmultitype, do2023_ComponentBehaviourExcess}. With more blocks and no intrablock edges, it becomes the multipartite Erdős–Rényi model, whose phase transition and giant-component behavior are covered by the more general theory of finite-type inhomogeneous random graphs~\cite{bollobas2005_phasetransitioninhomogeneous, soderberg2002_Generalformalisminhomogeneous}.

The second SBMS has a growing number of blocks, where the number of communities and the size of each community both grow as $\sqrt{n}$, while intracommunity and intercommunity probabilities decay at rates that give a finite limiting average degree.

The third SBMS is one in which the communities are the layers of a perfect $a$-ary tree, the number of communities grows with $n$, and edges are allowed only between consecutive layers. This sequence is motivated by the use of trees as a benchmark for percolation methods, since message-passing or self-consistency equations are exact on trees~\cite{allard2019_PercolationEffectiveStructure, karrer2014_Percolationsparsenetworks, newman2023_MessagePassingMethods, mann2025_AlternativeExpressionMessage}.

\begin{figure*}
    \centering
    \includegraphics[width=\linewidth, trim={0cm 0.3cm 0cm 0.25cm}, clip]{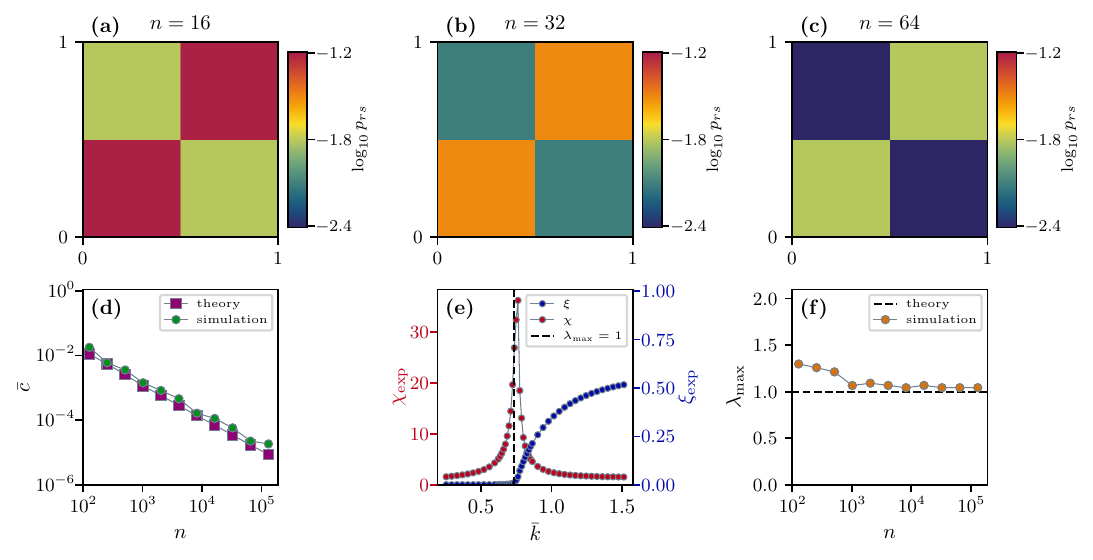}
    \caption{\textbf{SBM with a constant number of blocks.} (a)--(c) The connection-probability maps showing two blocks whose relative sizes remain fixed while all connection probabilities decrease as $n$ increases. (d) The average local clustering coefficient $\bar{c}$ at the experimentally determined critical point, showing both the theoretical prediction from Eq.~\eqref{eq:general-sbm-local-clustering-unconditioned} and the average value measured in simulations. (e) The experimental GCC membership probability $\xi_{\mathrm{exp}}$ and susceptibility $\chi_{\mathrm{exp}}$ (defined in Sec.~\ref{sec:numerics}) as functions of the average degree $\bar{k}$; the vertical line marks the value of $\bar{k}$ corresponding to the spectral threshold $\lambda_{\max} = 1$. (f) The largest eigenvalue $\lambda_{\max}$ evaluated at the experimentally determined critical point and compared with the theoretical prediction $\lambda_{\max}=1$.}
    \label{fig:const}
\end{figure*}

For each of these SBM sequences, we derive the spectral threshold predicted by the linearized self-consistency method, compute the corresponding  average degree, and compare that critical average degree with $1$.

\subsection{Constant number of blocks}\label{sec:constant-block-sbm}

We begin with the classical sparse SBM sequence with a fixed number of communities. We take $q(n)=q$, where $q$ is a fixed constant. The community sizes grow linearly as $n_r(n) = \alpha_r n$, where $\alpha_1, \ldots, \alpha_q$ are positive constants summing to $1$, and the connection probabilities scale as $p_{rs}(n) = \beta_{rs}/n$, where $\beta_{rs}$ are fixed connectivity constants. A sample sequence of connection-probability maps for this SBMS is shown in Fig.~\ref{fig:const}(a)--(c), illustrating that the number of blocks remains fixed, while all connection probabilities decay.

Because $q$ is fixed and every connection probability scales as $\mathcal{O}(1/n)$, each vertex has bounded expected degree. Consequently, for any fixed radius, the number of vertices in the neighborhood of a uniformly chosen vertex is bounded with high probability by a constant independent of $n$. Any cycle in this neighborhood requires an additional edge between two vertices already connected by a path. Since there are only a bounded number of possible such edges and each is present with probability $\mathcal{O}(1/n)$, the probability that the neighborhood contains a cycle tends to zero. Thus, the sequence is asymptotically locally tree-like, and the linearized self-consistency method from Sec.~\ref{sec:linearized-self-consistency} applies. As a result of this local tree-likeness, the probability that two neighbors of a vertex are connected tends to zero, so $\bar{c} \to 0$, as shown theoretically and numerically in Fig.~\ref{fig:const}(d).

For this SBMS, the diagonal terms of the linearized matrix are $t_{rr}(n) = (n_r-1)p_{rr}$, while the off-diagonal terms are given by $t_{rs}(n)=n_sp_{rs}$ for $s \neq r$. Then, substituting $n_r=\alpha_r n$ and $p_{rs}=\beta_{rs}/n$, we get
\begin{equation}\label{eq:linearized-matrix-const}
\begin{aligned}
    t_{rr}(n) &= (\alpha_rn-1)\frac{\beta_{rr}}{n} = \alpha_r \beta_{rr} - \frac{\beta_{rr}}{n}, \\
    t_{rs}(n) &= \alpha_s \beta_{rs} \quad (s \neq r).
\end{aligned}
\end{equation}
Therefore, in the limit $n \to \infty$, we see that the matrix $t(n)$ converges to the $q \times q$ matrix $t$ with entries
\begin{equation}\label{eq:linearized-matrix-const2}
    t_{rs} = \alpha_s \beta_{rs}.
\end{equation}
By Eq.~\eqref{eq:fixed-point-threshold}, the predicted asymptotic percolation threshold is therefore $\lambda_{\max}(t)=1$.

We next compare the predicted threshold with the average-degree condition $\bar{k}=1$. Using the general formula for $\bar{k}$ given in Eq.~\eqref{eq:general-sbm-average-degree}, we obtain
\begin{equation}\label{eq:const-blocks-average-degree}
\begin{aligned}
\bar{k} = \frac{1}{n}\sum_{r=1}^q \alpha_r n \left(\frac{\beta_{rr}}{n}(\alpha_r n - 1) + \sum_{s \neq r} \frac{\beta_{rs}}{n}\alpha_sn\right). 
\end{aligned}
\end{equation}
Hence, rewriting this and then taking $n \to \infty$, we arrive at
\begin{equation}\label{eq:const-blocks-limiting-average-degree}
\begin{aligned}
\bar{k} &= \sum_{r, s = 1}^q \alpha_r \alpha_s \beta_{rs} - \frac{1}{n}\sum_{r=1}^q \alpha_r \beta_{rr} \\
&\overset
{n \to \infty}{\longrightarrow} \sum_{r,s=1}^q \alpha_r \alpha_s \beta_{rs}. 
\end{aligned}
\end{equation}
The limiting average degree $\sum_{r, s=1}^q \alpha_r \alpha_s \beta_{rs}$ does not generally equal the largest eigenvalue of the matrix with entries $t_{rs}=\alpha_s \beta_{rs}$. Thus, in this fixed-block SBMS, the critical point is not generally characterized by $\bar{k}=1$.

\begin{figure}
    \centering
    \includegraphics[width=\linewidth, trim={7.5cm 1cm 6.75cm 0cm}, clip]{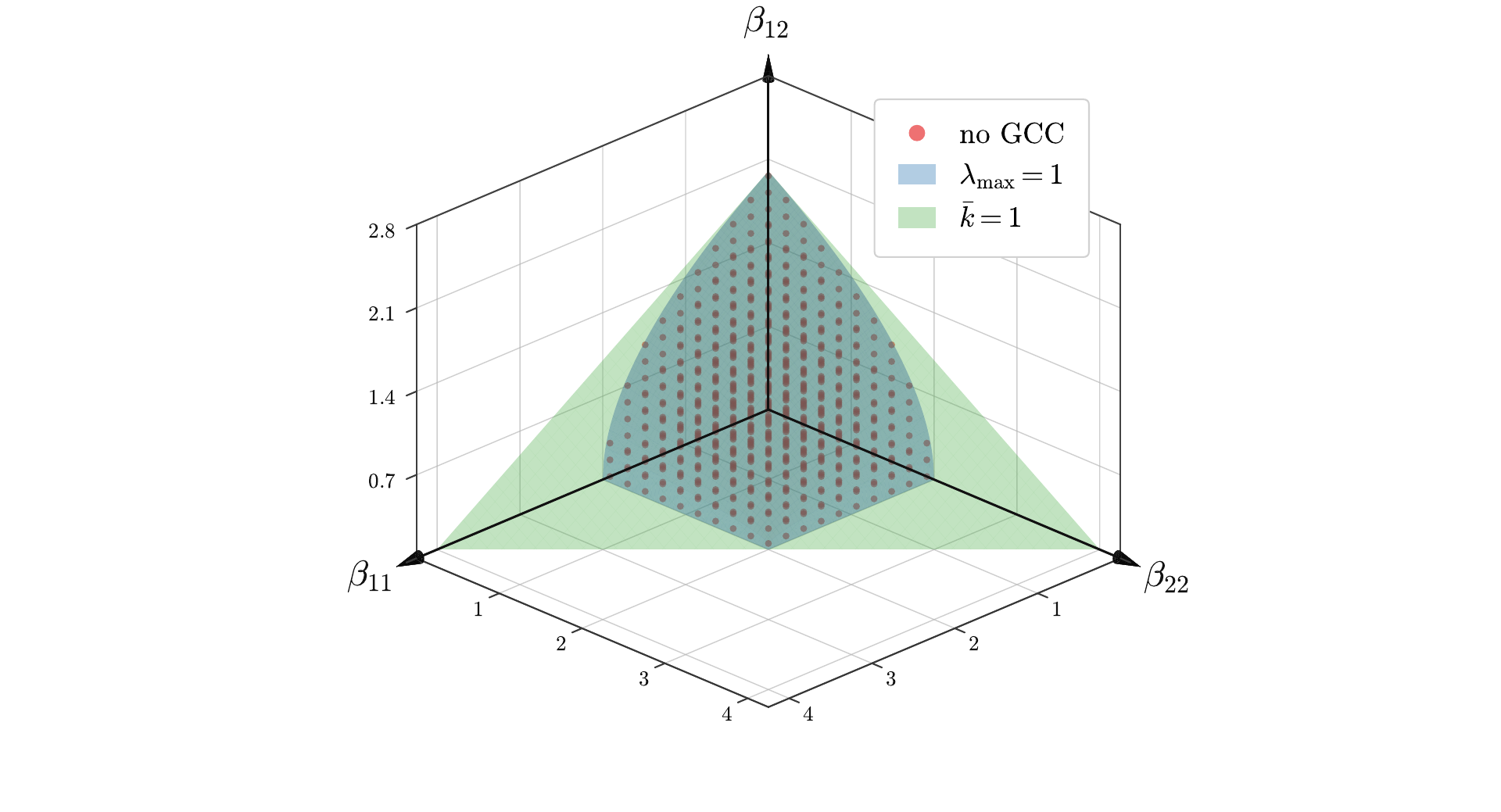}
    \caption{\textbf{Spectral versus average degree percolation conditions in the two-block SBM.} The points indicate parameter values for which the self-consistent equations \eqref{eq:fixed-point-xi} admit only the trivial solution $(\xi_1, \xi_2) = (0,0)$, corresponding to the absence of a GCC. Parameter values with a GCC are omitted so that the boundary of the no-GCC region remains visible. The blue surface shows the spectral threshold $\lambda_{\max}=1$, while the green surface shows the average-degree condition $\bar{k}=1$.}
    \label{fig:parameter-space}
\end{figure}

To illustrate this difference, we compare the conditions $\lambda_{\max}=1$ and $\bar{k}=1$ across the full parameter space with $q=2$. We evaluate a $21 \times 21 \times 21$ grid of parameter triples $(\beta_{11}, \beta_{12}, \beta_{22})$, with $\beta_{11}, \beta_{22} \in [0,4]$ and $\beta_{12}=\beta_{21} \in [0, 2.8]$. For each parameter triple, we numerically solve the self-consistent equations \eqref{eq:fixed-point-xi} to determine whether a GCC exists. As shown in Fig.~\ref{fig:parameter-space}, the boundary of the region without a GCC follows $\lambda_{\max}=1$, rather than $\bar{k}=1$.

Although the conditions $\lambda_{\max}=1$ and $\bar{k}=1$ differ in general, they coincide in certain special cases. In bipartite Erdős–Rényi graphs, $q=2$, $\beta_{11}=\beta_{22}=0$, $\beta_{12}=\beta_{21}=\beta$, and if the two blocks are of the same size, $\alpha_1=\alpha_2=1/2$. Hence, $\lambda_{\max} = \beta/2 = \bar{k}$, so $\lambda_{\max}(t) = 1$ is equivalent to $\bar{k}=1$. Our criterion therefore recovers the rigorously established bipartite Erdős–Rényi transition~\cite{johansson2012_giantcomponentrandom, do2023_ComponentBehaviourExcess}.

We now check our analytical results in simulations. We take two communities of equal size, $\alpha_1 = \alpha_2 = 1/2$, and increase the number of vertices in powers of $2$ from $n=128$ to $n=131{,}072$. We fix the connectivity parameters $\beta_{11}=0.5$ and $\beta_{12}=\beta_{21}=0.25$, and vary $\beta_{22}\in [0.01, 5.0]$ as the control parameter. Figure~\ref{fig:const}(e) shows the average relative size of the largest connected component, $\xi_{\mathrm{exp}}$, and the susceptibility, $\chi_\mathrm{exp}$, as functions of the average degree $\bar{k}$ for graphs of size $n=131{,}072$, together with the theoretical threshold $\lambda_{\max}=1$.

For this numerical example, $t(n)$ is a symmetric $2 \times 2$ matrix, so its largest eigenvalue can be written in closed form as
\begin{equation}\label{eq:exact-2x2-lambda-max}
    \lambda_{\max}(t(n))
    =
    \frac{t_{11} + t_{22}}{2}
    +
    \sqrt{
        \left( \frac{t_{11} - t_{22}}{2} \right)^2
        +
        t_{12}^2
    }.
\end{equation}
Evaluating Eq.~\eqref{eq:exact-2x2-lambda-max} at the experimentally determined critical value $\beta_c$, we find that $\lambda_{\max}$ approaches the theoretical value $1$ as $n$ increases, as shown in Fig.~\ref{fig:const}(f).

\subsection{Growing number of blocks}\label{sec:growing-block-sbm}

\begin{figure*}[tb]
    \centering
    \includegraphics[width=\linewidth, trim={0cm 0.3cm 0cm 0.25cm}, clip]{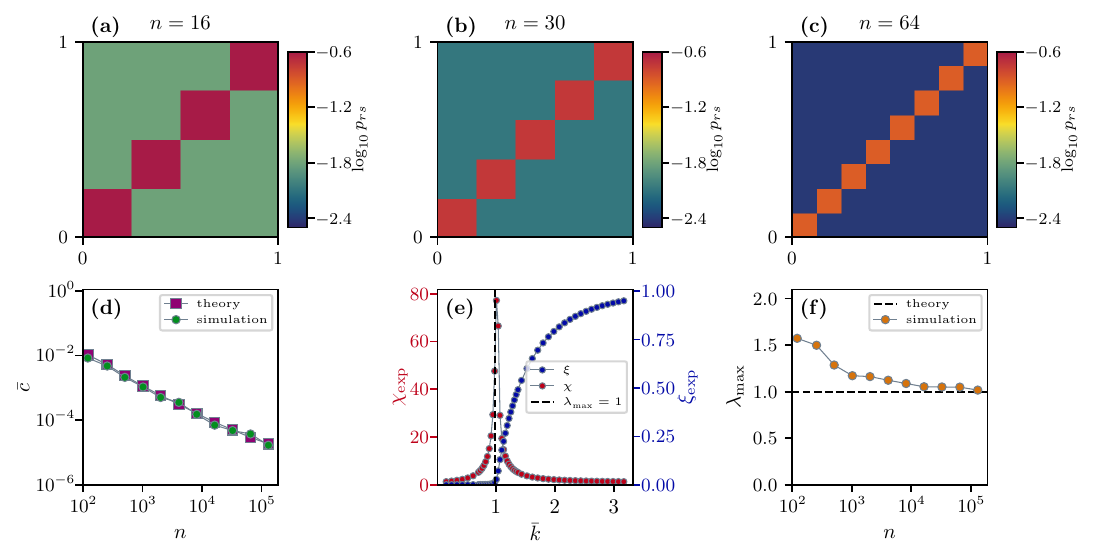}
    \caption{\textbf{SBM with a growing number of blocks.} (a)--(c) The connection-probability maps for a sequence with an increasing number of equal-size blocks and stronger intrablock than interblock connectivity. (d) The average local clustering coefficient $\bar{c}$ evaluated at the experimentally determined critical point. The theoretical prediction from Eq.~\eqref{eq:general-sbm-local-clustering-unconditioned} is compared with the average value measured from simulations, with both tending to zero as $n$ increases. (e) The experimental GCC membership probability $\xi_{\mathrm{exp}}$ and susceptibility $\chi_{\mathrm{exp}}$ (defined in Sec.~\ref{sec:numerics}) as functions of the average degree $\bar{k}$; the vertical line indicates the average degree corresponding to the spectral threshold $\lambda_{max}=1$, which coincides with $\bar{k}=1$ in this SBMS. (f) The largest eigenvalue $\lambda_{max}$ at the experimentally determined critical point, demonstrating agreement with the theoretical prediction $\lambda_{\max}=1$.}
    \label{fig:sqrt}
\end{figure*}

Next, we consider an SBM sequence in which both the number of communities and the size of each community grow as $\sqrt{n}$. We consider graph sizes $n=m^2$, where $m \to \infty$, and take $q(n)=\sqrt{n}$ communities, each of size $n_r(n) = \sqrt{n}$. The intrablock probabilities decay as $p_{rr}(n) = a/\sqrt{n}$ for some fixed constant $a>0$, while the interblock probabilities decay as $p_{rs}(n)=b/n$ for $r \neq s$, for some fixed constant $b>0$. A sample sequence of connection-probability maps for this SBMS is shown in Fig.~\ref{fig:sqrt}(a)--(c), illustrating that the number of blocks grows with $n$, while the larger intrablock probabilities remain concentrated along the diagonal.

Although the number of communities grows, the neighborhood of any vertex remains sparse. Specifically, within its own community, a vertex has on the order of $\sqrt{n}$ possible neighbors, each connected with probability on the order of $1/\sqrt{n}$; outside its own community, it has on the order of $n$ possible neighbors, each connected with probability on the order of $1/n$. Its expected degree therefore remains bounded, as does the number of vertices reached within any fixed distance with high probability. For a cycle to occur in such a neighborhood, at least one additional connection must be present among this bounded number of vertices, so since the largest connection probability is $\max\{a/\sqrt{n}, b/n \} \to 0$, the probability of such a cycle tends to zero. The sequence is therefore asymptotically locally tree-like, allowing the linearized self-consistency method from Sec.~\ref{sec:linearized-self-consistency} to be applied. The corresponding disappearance of triangles is reflected in $\bar{c} \to 0$, as shown theoretically and numerically in Fig.~\ref{fig:sqrt}(d).

The diagonal terms of $t(n)$ are $t_{rr}(n)=(n_r-1)p_{rr}$, while its off-diagonal terms are $t_{rs}(n)=n_sp_{rs}$ for $s \neq r$. Substituting $n_r=\sqrt{n}, p_{rr}=a/\sqrt{n}$, and $p_{rs}=b/n$, we obtain
\begin{equation}\label{eq:linearized-matrix-sqrt}
\begin{aligned}
    t_{rr}(n) &= \left(\sqrt{n}-1\right)\frac{a}{\sqrt{n}} = a - \frac{a}{\sqrt{n}}, \\
    t_{rs}(n) &= \left(\sqrt{n}\right)\frac{b}{n} = \frac{b}{\sqrt{n}} \quad (s \neq r).
\end{aligned}
\end{equation}
Since this is a $\sqrt{n} \times \sqrt{n}$ matrix with diagonal entries $a-a/\sqrt{n}$ and off-diagonal entries $b/\sqrt{n}$, the largest eigenvalue is therefore just the row sum,
\begin{equation}\label{eq:largest-eigenvalue-sqrt}
\begin{aligned}
    \lambda_{\max}\left(t(n)\right) &= a - \frac{a}{\sqrt{n}} + \left(\sqrt{n}-1\right)\frac{b}{\sqrt{n}} \\
    &= a+b - \frac{a+b}{\sqrt{n}}.
\end{aligned}
\end{equation}
Taking $n \to \infty$, we obtain $\lambda_{\max}(t(n)) \to a+b$, so the loss of stability of the trivial fixed point $\boldsymbol{\xi} = \mathbf{0}$ of the self-consistent equation~\eqref{eq:self-consistent-equation} occurs at $a+b=1$. Because all communities are statistically equivalent, $\xi_1 = \cdots = \xi_q$. Hence, any nontrivial solution $\boldsymbol{\xi}$ gives
\begin{equation}
    \xi = \sum_{r=1}^q \frac{n_r}{n} \xi_r = \xi_1 > 0.
\end{equation}
Thus, the loss of stability indeed corresponds to a nonzero GCC membership probability, so the predicted transition is also at $a+b=1$.

We next determine whether the threshold condition $\lambda_{\max}=1$ agrees with the average-degree condition $\bar{k}=1$. Since all the communities have the same size and connection probabilities, Eq.~\eqref{eq:general-sbm-average-degree} gives
\begin{equation}\label{eq:sqrt-average-degree}
    \bar{k} = \frac{1}{n}\sum_{r=1}^{\sqrt{n}} \sqrt{n}\left(\frac{a}{\sqrt{n}}\left(\sqrt{n}-1\right)+\sum_{s \neq r} \frac{b}{n}\left(\sqrt{n}\right)\right).
\end{equation}
Since there are $\sqrt{n}-1$ terms in the interblock sum, this becomes
\begin{equation}\label{eq:sqrt-average-degree-limiting}
\begin{aligned}
    \bar{k} &= \left(a - \frac{a}{\sqrt{n}}\right) + \left(b - \frac{b}{\sqrt{n}}\right) \\
    &= a+b - \frac{a+b}{\sqrt{n}} \\
    &\overset{n \to \infty}{\longrightarrow} a+b.
\end{aligned}
\end{equation}
Comparing Eqs. \eqref{eq:largest-eigenvalue-sqrt} and \eqref{eq:sqrt-average-degree-limiting}, we obtain $\lambda_{\max}(t(n))=\bar{k}$ for every $n$, and both quantities converge to $a+b$ as $n \to \infty$. Hence, the predicted transition at $\lambda_{\max}=1$ occurs at $\bar{k}=1$.

\begin{figure*}
    \centering
    \includegraphics[width=\linewidth, trim={0cm 0.3cm 0cm 0.25cm}, clip]{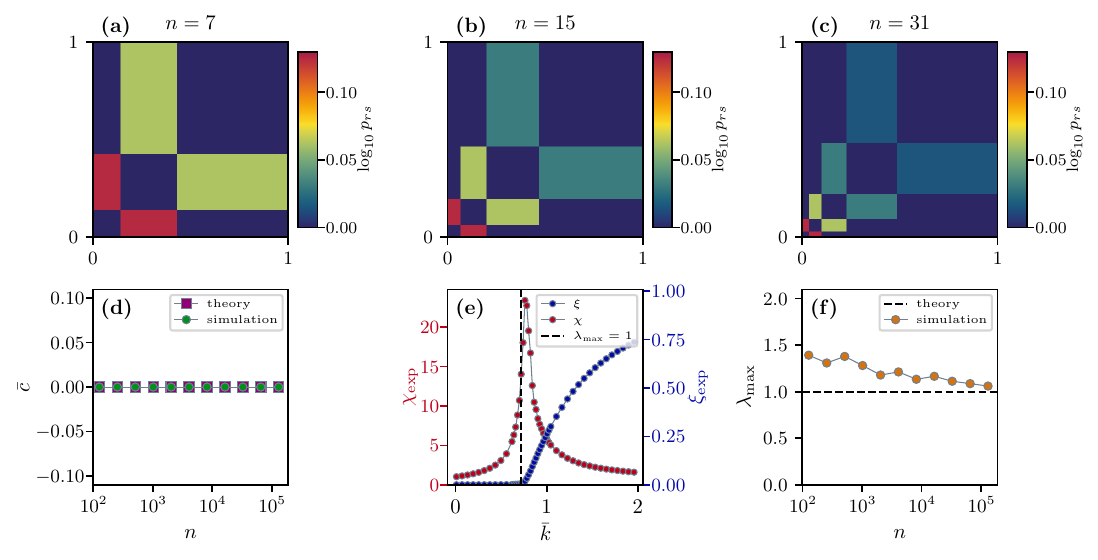}
    \caption{\textbf{SBM with perfect $a$-ary tree layers as blocks.} (a)--(c) The connection-probability maps for communities whose sizes follow the layers of a perfect $a$-ary tree, with edges only between consecutive layers. (d) The average local clustering coefficient $\bar{c}$ evaluated at the experimentally determined critical point. Both the theoretically predicted average clustering and the average value of clustering measured from simulations are identically zero because this SBMS contains no triangles.  (e) The experimental GCC membership probability $\xi_{\mathrm{exp}}$ and susceptibility $\chi_{\mathrm{exp}}$ (defined in Sec.~\ref{sec:numerics}) as functions of the average degree $\bar{k}$; the vertical line indicates the value of $\bar{k}$ at the spectral threshold $\lambda_{\max}=1$. (f) The largest eigenvalue $\lambda_{\max}$ at the experimentally determined critical point, demonstrating agreement with the theoretical prediction $\lambda_{\max}=1$.}
    \label{fig:tree}
\end{figure*}

The predicted percolation threshold is compared with simulations using $n = \lfloor \sqrt{n'} \rfloor^2$ with $n'$ increasing in powers of $2$ from $n'=128$ to $n'=131{,}072$. We fix the intrablock connectivity parameter $a=0.15$ and vary the interblock connectivity as a control parameter, $b \in [0.01, 3.0]$. Figure~\ref{fig:sqrt}(e) shows the average relative size of the largest connected component, $\xi_\mathrm{exp}$, and the susceptibility, $\chi_\mathrm{exp}$, as functions of $\bar{k}$ for graphs of size $n=\lfloor \sqrt{131{,}072}\rfloor^2$, together with the common threshold $\lambda_{\max}(t(n))=\bar{k}=1$. Evaluating Eq.~\eqref{eq:largest-eigenvalue-sqrt} at the experimentally determined critical point $b_c$, we find that $\lambda_{\max}(t(n))$ approaches $1$ as $n$ increases, as shown in Fig.~\ref{fig:sqrt}(f).

\subsection{Perfect $a$-ary tree layer SBM sequence}\label{sec:tree-layers-sbm}

We now consider an SBM sequence whose community sizes grow like the size of the layers of a perfect $a$-ary tree, where $a \ge 2$ is a fixed integer branching factor. For each $n$ in this sequence, let $\ell(n)$ be the height of this perfect $a$-ary tree, so that $q(n) = \ell(n)+1$. Let us index the communities by $0, 1, \ldots, \ell(n)$ based on the number of layers away from the root vertex, and we let the size of community $r$ be $n_r(n) = a^r$ for all $0 \le r \le \ell(n)$. Thus, $n = 1 + a + \cdots + a^{\ell(n)}$. Edges are only allowed between consecutive layers, i.e., $p_{rs}(n) = 0$ unless $|r-s|=1$, and for $0 \le r \le \ell(n)-1$ we set $p_{r, r+1}(n) = p_{r+1, r}(n) = b/(a^{r+1})$. Although the community sizes match the layer sizes of a perfect $a$-ary tree, sampled graphs are generally not perfect $a$-ary trees. Indeed, each vertex has an expected $b$ neighbors in the next layer and $b/a$ neighbors in the preceding layer, rather than being connected to exactly $a$ neighbors in the next layer and $1$ neighbor in the previous layer. A sample sequence of connection-probability maps for this SBMS is shown in Fig.~\ref{fig:tree}(a)--(c), illustrating that the number of layers grows with $n$, while edges are restricted to adjacent layers only.

To establish local tree-likeness, consider a uniformly chosen vertex and let $r$ be the layer containing it. For every fixed $r_0$,
\begin{equation}
    \text{Pr}(r \le r_0) = \frac{1+a+\cdots + a^{r_0}}{1+a+\cdots+a^{\ell(n)}} \to 0.
\end{equation}
Thus, the layer containing a uniformly chosen vertex moves increasingly far from the innermost layers as $n \to \infty$. The expected neighbor counts given above show that every vertex has bounded expected degree, so the number of vertices reached within any fixed distance is bounded with high probability. Such a neighborhood reaches only layers whose indices differ from $r$ by a bounded amount, and the connection probabilities between these layers tend to zero, so since a cycle would require an additional edge among the bounded number of vertices in the neighborhood, the probability of a cycle tends to zero. Thus, the sequence is asymptotically locally tree-like, so the linearized self-consistency method from Sec.~\ref{sec:linearized-self-consistency} applies. Because edges occur only between consecutive layers, the model also contains no triangles, giving $\bar{c}=0$ for every $n$, as confirmed by the theoretical and simulated values in Fig.~\ref{fig:tree}(d).

Since there are no intralayer edges, all diagonal entries of $t(n)$ are zero. For consecutive layers, the off-diagonal entries are
\begin{equation}\label{eq:linearized-matrix-tree}
\begin{aligned}
    t_{r, r+1}(n) &= n_{r+1}p_{r, r+1} = a^{r+1} \cdot \frac{b}{a^{r+1}} = b, \\
    t_{r+1, r}(n) &= n_r p_{r+1, r} = a^r \cdot\frac{b}{a^{r+1}} = \frac{b}{a}.
\end{aligned}
\end{equation}
Therefore, $t(n)$ is a tridiagonal $(\ell(n)+1) \times (\ell(n)+1)$ matrix with upper diagonal entries $b$ and lower diagonal entries $b/a$. This matrix is similar to the symmetric Toeplitz tridiagonal matrix with all diagonal entries equal to $0$ and all nonzero off-diagonal entries equal to $b/\sqrt{a}$. Applying the standard eigenvalue formula for symmetric tridiagonal Toeplitz matrices \cite{meurant2025_HessenbergTridiagonalMatrices}, we obtain the eigenvalues
\begin{equation}\label{eq:linearized-matrix-tree-eigenvalues}
    \lambda_j \left(t(n)\right) = \frac{2b}{\sqrt{a}}\cos\left(\frac{j\pi}{\ell(n)+2}\right),
\end{equation}
for $j=1, 2, \ldots, \ell(n)+1$. Hence, it follows that
\begin{equation}\label{eq:largest-eigenvalue-tree}
\begin{aligned}
    &\qquad \lambda_{\max} \left(t(n)\right) = \frac{2b}{\sqrt{a}}\cos\left(\frac{\pi}{\ell(n)+2}\right) \\
    &\implies b_c = \frac{\sqrt{a}}{2\cos\left(\tfrac{\pi}{\ell(n)+2}\right)} \overset{n \to \infty}{\longrightarrow} \frac{\sqrt{a}}{2},
\end{aligned}
\end{equation}
so the predicted percolation threshold is given by $b_c = \tfrac{\sqrt{a}}{2}$ in the $n \to \infty$ limit.

Because the number of layers grows with $n$, we must also verify that the instability is not confined to the innermost layers. Let $t'$ be the $h \times h$ linearized matrix for the subgraph induced by the $h$ outermost layers. Then, the largest eigenvalue of this matrix is
\begin{equation}
    \lambda_{\max}(t') = \frac{2b}{\sqrt{a}}\cos\left(\frac{\pi}{h+1}\right).
\end{equation}
If $b > \sqrt{a}/2$, then $h$ can be chosen sufficiently large, but fixed independently of $n$, so that this eigenvalue is greater than $1$. Moreover, the fraction of vertices in these $h$ outermost layers satisfies
\begin{equation}
    \frac{a^{\ell(n)-h+1}+\cdots+a^{\ell(n)}}{1+a+\cdots+a^{\ell(n)}} \longrightarrow 1 - a^{-h} > 0.
\end{equation}
The SBM induced by these layers is therefore supercritical and occupies a positive fraction of the full graph. Thus, the instability above $b_c=\sqrt{a}/2$ indeed produces a GCC whose relative size remains positive as $n \to \infty$.

We next compute the average degree corresponding to this threshold. The expected number of edges between layers $r$ and $r+1$ is $n_r n_{r+1} p_{r, r+1} = a^r  \cdot a^{r+1} \cdot b/(a^{r+1}) = ba^r$, so the expected total number of edges is
\begin{equation}
    \sum_{r=0}^{\ell(n)-1} ba^r = \frac{b(a^{\ell(n)}-1)}{a-1}.
\end{equation}
The average degree is equal to twice the expected total number of edges divided by the number of vertices, giving
\begin{equation}\label{eq:tree-average-degree-limiting}
\begin{aligned}
    \bar{k} &= \frac{2 \cdot \tfrac{b(a^{\ell(n)}-1)}{a-1}}{\tfrac{a^{\ell(n)+1}-1}{a-1}} \\
    &\overset{n \to \infty}{\longrightarrow} \frac{2b}{a}.
\end{aligned}
\end{equation}
At the limiting threshold $b_c = \sqrt{a}/2$, Eq.~\eqref{eq:tree-average-degree-limiting} gives
\begin{equation}\label{eq:critical-average-degree-tree}
    \bar{k}_c = \frac{1}{\sqrt{a}.}
\end{equation}
Since $a \ge 2$, it follows that $\bar{k}_c < 1$. Thus, the GCC in this SBMS emerges before the average degree reaches $1$.

For numerical evaluation, we fix $a=2$ and take $q(n) \in \{7,8,\dots, 17\}$, which determines the graph size $n$. We vary $b \in [0.01, 2]$ as the control parameter. Fig.~\ref{fig:tree}(e) shows the average relative size of the largest connected component, $\xi_{\mathrm{exp}}$, and the susceptibility, $\chi_{\mathrm{exp}}$, as functions of $\bar{k}$, together with the vertical line marking the value of $\bar{k}$ corresponding to $\lambda_{\max}=1$. Evaluating Eq.~\eqref{eq:largest-eigenvalue-tree} at the experimentally determined critical value $b_c$, we find that $\lambda_{\max}(t(n))$ approaches $1$ as $n$ increases, as shown in Fig.~\ref{fig:tree}(f).

\section{Clustered SBM sequences}\label{sec:nonzero-clustering-sbms}

We now consider two SBM sequences in which the neighborhood of a uniformly chosen vertex contains short cycles with nonvanishing probability as $n \to \infty$. These sequences are therefore not asymptotically locally tree-like, so the independence assumption underlying Eq.~\eqref{eq:fixed-point-xi} fails: different paths from a vertex can pass through the same local structure, and the corresponding failure probabilities cannot be multiplied independently. The linearized self-consistency method from Sec.~\ref{sec:linearized-self-consistency} therefore does not provide an accurate prediction of the percolation threshold. This persistent local cyclic structure is also reflected in the nonvanishing limit of $\bar{c}$. We instead turn to the community-scale branching process method from Sec.~\ref{sec:community-branching-process}, which applies directly to the first sequence and is used as an approximation for the second.

The two SBM sequences in this section produce this clustering through different mechanisms. The first is an Erdős–Rényi SBMS with dense planted communities, where the number of communities grows linearly with $n$, each community has fixed size, and the intracommunity edge probability remains constant. This planted-partition form is standard in work on community detection and its statistical and computational transitions~\cite{karrer2011_StochasticBlockmodelsCommunity, abbe2018_CommunityDetectionStochastic, decelle2011_AsymptoticAnalysisSBM}, and has more recently been studied in connection with the overlap gap property of modularity~\cite{bhamidi2026_StochasticBlockModel}. Our scaling differs from the standard fixed-block regime, however, because the number of communities grows while their sizes remain fixed. For percolation, the closest comparison is the classical blockmodel, whose threshold and giant-component behavior have been studied in~\cite{bujok2014_PercolationClassicalBlockmodel}.

The second SBMS is geometric, in which fixed-size communities are arranged on a circle and their connection probabilities are induced by a distance-dependent geometric kernel. This construction connects the model to continuum percolation and soft random geometric graphs, for which percolation, giant-component behavior, and connectivity have been studied~\cite{penrose1991_continuumpercolationmodel, penrose2022_Giantcomponentsoft, dubin2026_GeometryGiantComponent, penrose2016_ConnectivitySoftRandom}. It is also related to latent-space network models, in which edge probabilities depend on positions in an underlying geometric space~\cite{hoff2002_LatentSpaceApproaches}. While the geometric SBMS is not projective for finite $n$~\cite{spencer2023_Projectivesparselearnable}, it is noteworthy that in the thermodynamic limit the cyclic wrap-around becomes negligible and the model locally approaches the growing interval model on an infinite line, which \emph{is} projective.

\subsection{Erdős–Rényi with dense planted communities}\label{sec:dense-community-sbm}

\begin{figure*}
    \centering
    \includegraphics[width=\linewidth, trim={0cm 0.3cm 0cm 0.25cm}, clip]{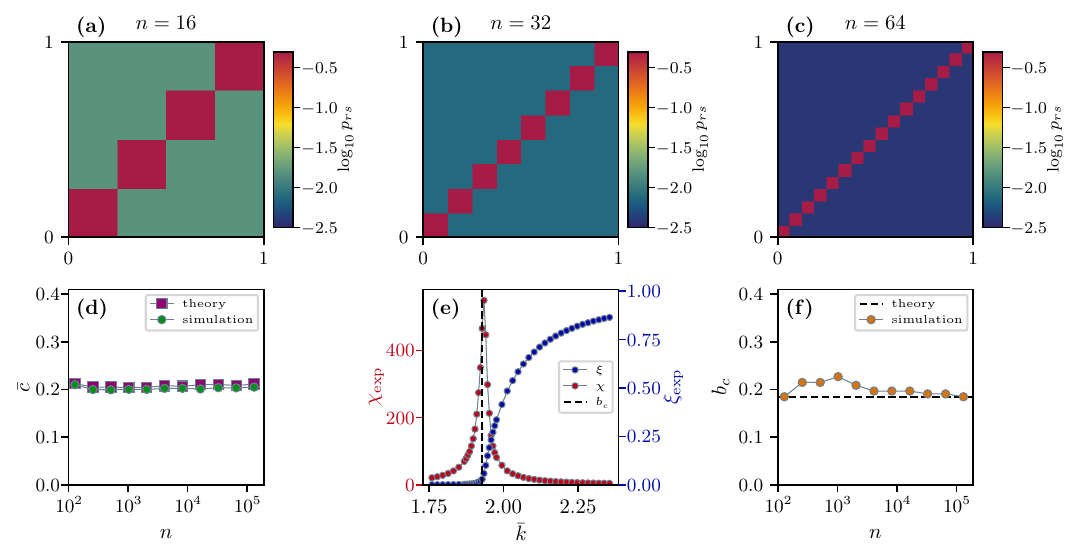}
    \caption{\textbf{Erdős–Rényi with dense planted communities.} (a)--(c) The connection-probability maps for a sequence with a growing number of fixed-size dense communities connected by sparse intercommunity edges. (d) The average local clustering coefficient $\bar{c}$ evaluated at the experimentally determined critical point. The theoretical prediction from Eq.~\eqref{eq:general-sbm-local-clustering-unconditioned}  is compared with the average value measured from simulations, with both converging to the same nonzero limit as $n$ increases. (e) The experimental GCC membership probability $\xi_{\mathrm{exp}}$ and susceptibility $\chi_{\mathrm{exp}}$ (defined in Sec.~\ref{sec:numerics}) as functions of the average degree $\bar{k}$; the vertical line indicates the average degree at the critical value of $b$, i.e., $\mathbb{E}[T]b_c=1$.  (f) The experimentally determined critical value of $b$ compared with the corresponding theoretical prediction $b_c=1/\mathbb{E}[T]$.}
    \label{fig:linear}
\end{figure*}

We first consider an SBM sequence with dense planted communities. Let $s \ge 3$ be fixed, and let the number of communities $q(n)$ grow as $n/s$, with each community of size $n_r(n)=s$. The intracommunity probabilities are fixed at $p_{rr}(n)=p$, where $0<p<1$, while the intercommunity probabilities decay as $p_{rt}(n)=b/n$ for $r \neq t$, where $b>0$ is a constant. A sample sequence of connection-probability maps for this SBMS is shown in Fig.~\ref{fig:linear}(a)--(c), illustrating that the number of communities grows linearly with $n$, while the dense intracommunity probabilities remain concentrated along the diagonal.

Here, both the community size $s$ and the intracommunity probability $p$ remain fixed. Since $s \ge 3$, consider a vertex and two other vertices in its community. The probability that these three vertices form a triangle is $p^3 > 0$, independently of $n$. Thus, short cycles persist in the neighborhoods of vertices, and the sequence is not locally tree-like. The same triangles also produce a nonzero limiting average local clustering coefficient. Figure~\ref{fig:linear}(d) compares the clustering measured at the numerically determined critical point with the estimate from Eq.~\eqref{eq:general-sbm-local-clustering-unconditioned}, while Appendix~\ref{app:clustering-derivations} verifies that $\lim_{n \to \infty} \bar{c}_n > 0$. The intercommunity probabilities decay as $b/n$, however, so the sparse connections between communities still allow us to use the community-scale branching process from Sec.~\ref{sec:community-branching-process} to find the percolation threshold.

We begin by ignoring all intercommunity edges and fixing a vertex $v$. Let $T$ be the size of the connected component containing $v$ inside its own community. Since the intracommunity graph is $G(s,p)$, the distribution of $T$ is the distribution of the size of the component containing a fixed vertex in $G(s,p)$. Using the classical recurrence for the connectedness probabilities of finite Erdős–Rényi graphs~[\citenum{gilbert1959_RandomGraphs},
 \citenum{jiang2025_RobustnessSmallNetworks}], we obtain
\begin{equation}\label{eq:expected-size-connected-component-linear}
    \mathbb{E}[T] = \sum_{k=1}^s k\binom{s-1}{k-1} (1-p)^{k(s-k)} f_k,
\end{equation}
where $f_k$ is the probability that $G(k, p)$ is connected, given recursively by $f_1=1$ and
\begin{equation}\label{eq:prob-actually-connected-linear}
    f_k = 1 - \sum_{j=1}^{k-1} \binom{k-1}{j-1}(1-p)^{j(k-j)} f_j.
\end{equation}

Next, we compute the expected number $\bar{k}_{\mathrm{out}}$ of intercommunity edges leaving a single vertex. Since each vertex has $n-s$ possible neighbors outside of its own community, and each such edge is present with probability $b/n$, we have $\bar{k}_{\mathrm{out}} = (n-s)b/n = b(1-s/n) \overset{n \to \infty}{\longrightarrow} b$. Therefore, by the community-scale branching process condition from Sec.~\ref{sec:community-branching-process}, we arrive at the percolation threshold
\begin{equation}\label{eq:linear-percolation-threshold}
    \mathbb{E}[T]b = 1 \implies b_c = \frac{1}{\mathbb{E}[T]},
\end{equation}
where $\mathbb{E}[T]$ is a constant depending on $s$ and $p$ computed above.

We next determine the average degree at the predicted threshold. A vertex has $s-1$ possible neighbors within its own community, each joined to it with probability $p$, giving expected intracommunity degree $p(s-1)$. It has $n-s$ possible neighbors outside its community, each joined to it with probability $b/n$, giving expected intercommunity degree $(n-s)b/n$. Therefore,
\begin{equation}\label{eq:linear-average-degree-limiting}
\begin{aligned}
    \bar{k} &= p(s-1) + (n-s)\frac{b}{n} \\
    &= p(s-1) + b\left(1-\frac{s}{n}\right) \\
    &\overset{n \to \infty}{\longrightarrow} p(s-1)+b.
\end{aligned}
\end{equation}
At the percolation threshold, we have $b_c=1/\mathbb{E}[T]$, so
\begin{equation}\label{eq:linear-critical-average-degree}
    \bar{k}_c = p(s-1) + \frac{1}{\mathbb{E}[T]},
\end{equation}
which is not generally equal to $1$.

The prediction $b_c = 1/\mathbb{E}[T]$ is evaluated numerically using communities of size $s=8$. We take $n$ in powers of $2$ from $128$ to $131{,}072$, set $q(n)=n/8$, fix $p=0.25$, and vary $b \in [0.01, 0.6]$. Figure~\ref{fig:linear}(e) shows the relative size of the largest connected component, $\xi_{\mathrm{exp}}$, and the susceptibility, $\chi_{\mathrm{exp}}$, as functions of $\bar{k}$ for graphs of size $n=131{,}072$, with the average degree corresponding to $b_c = 1/\mathbb{E}[T]$ indicated. As $n$ increases, the experimentally determined critical value $b_c$ approaches $1/\mathbb{E}[T]$, as shown in Fig.~\ref{fig:linear}(f).

\subsection{Geometric SBM sequence}\label{sec:geometric-sbm}

\begin{figure}
    \centering
    \includegraphics[
        width=\linewidth,
        trim=10cm 5.4cm 10cm 3.15cm,
        clip
    ]{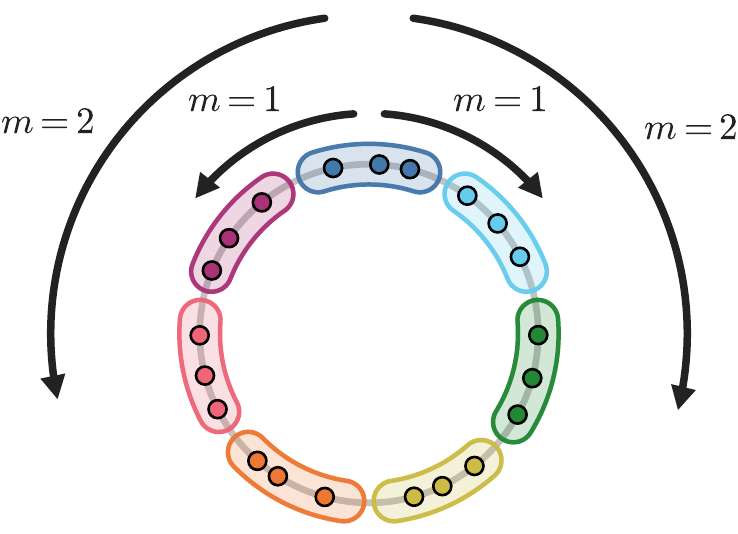}
    \caption{\textbf{Cyclic block distance in the geometric SBM.} Starting from a fixed reference community, the two nearest communities have cyclic block-distance $m=1$, the next two have cyclic block-distance $m=2$, and so on. The connection probability $\bar{p}(m)$ between two communities at cyclic block-distance $m$ is obtained by averaging the geometric kernel over all pairs of vertices in the corresponding pair of communities. By construction, $\bar{p}(m)$ depends only on the cyclic block-distance $m$, yielding the circular block structure of the geometric SBM.}
    \label{fig:gsbm-schematic}
\end{figure}

The final SBMS we consider is a geometric SBM sequence. In this SBMS, we approximate a soft random geometric graph on $[0, n)$ with periodic boundary conditions by an SBM sequence with $q(n)=n/s$ communities, each of fixed size $n_r(n)=s$. The communities are arranged cyclically around the ring, and the connection probabilities are induced by the geometric kernel
\begin{equation}\label{eq:geom-kernel}
    p(x) = \min \left\{ 1, \left(\frac{\rho}{x}\right)^{\beta}\right\},
\end{equation}
where $x$ denotes the wrapped distance around the circle, $\rho$ is the distance below which vertices are connected with probability $1$, and $1 < \beta < 2$ controls how quickly the connection probability decays for larger distances. Also, we impose the restriction $\rho<s$ so that the distance range where $p(x)=1$ is shorter than one block length. The connection probability between two communities is obtained by averaging $p(x)$ over all pairs of points in the two corresponding intervals. If two communities have cyclic block-distance $m$, as illustrated in Fig.~\ref{fig:gsbm-schematic}, we let their connection probability be denoted by $\bar{p}(m)$.

\begin{figure*}[p]
    \centering
    \includegraphics[width=\linewidth, trim={0cm 0.3cm 0cm 0.25cm}, clip]{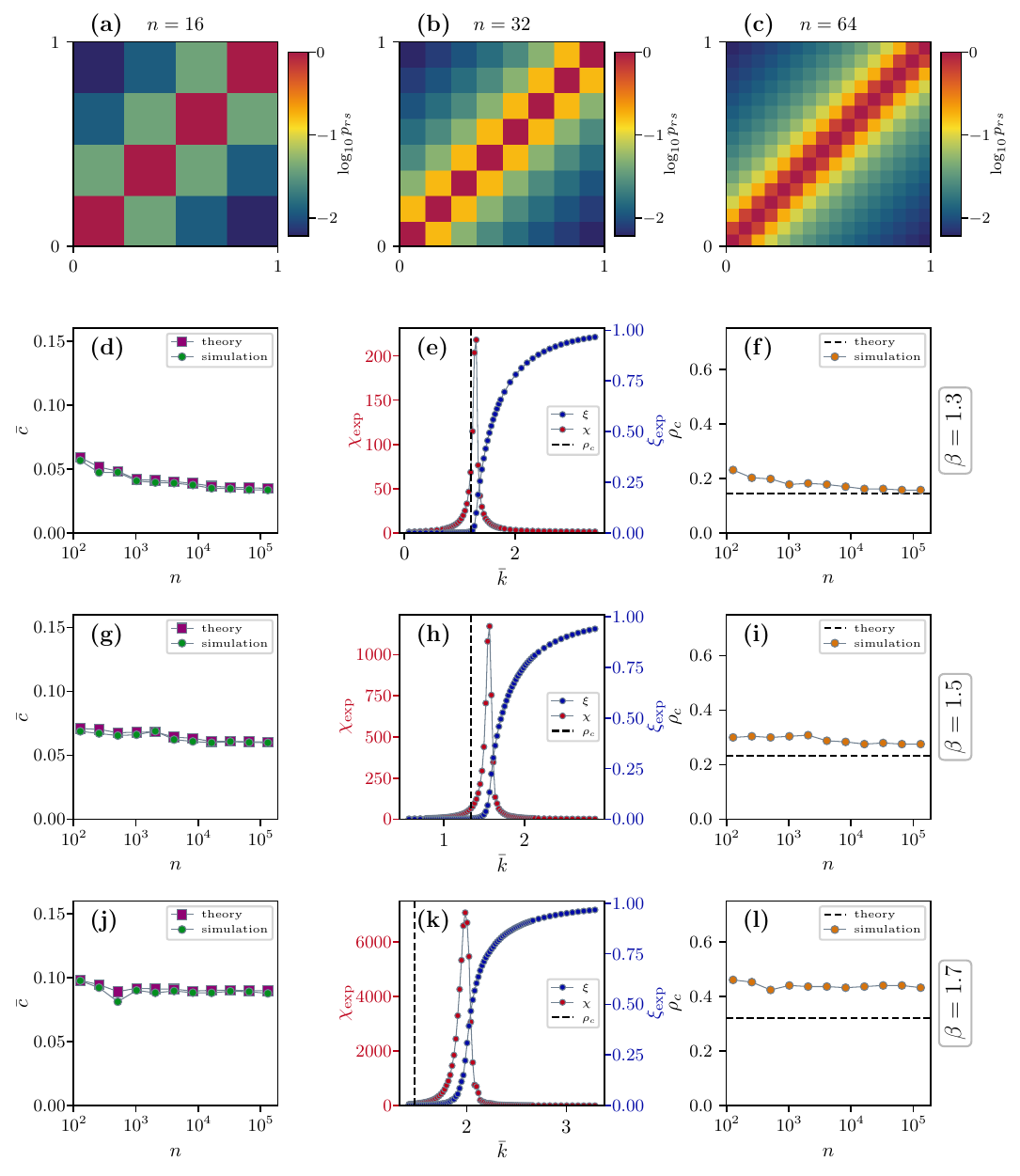}
    \caption{\textbf{Geometric SBMS.} (a)--(c) The connection-probability maps for communities arranged on a circle, with larger connection probabilities between nearby communities and smaller probabilities between more distant ones. Rows (d)--(f), (g)--(i), and (j)--(l) correspond to $\beta=1.3$, $\beta=1.5$, and $\beta=1.7$, respectively. In the first column, the theoretical average local clustering coefficient $\bar{c}$, obtained from Eq.~\eqref{eq:general-sbm-local-clustering-unconditioned}, is compared with the average value obtained from simulations, showing that clustering remains nonzero as $n$ increases. In the second column, the experimental GCC membership probability $\xi_{\mathrm{exp}}$ and susceptibility $\chi_{\mathrm{exp}}$ (defined in Sec.~\ref{sec:numerics}) are shown as functions of the average degree $\bar{k}$; the vertical lines mark the value of $\bar{k}$ corresponding to the critical value of $\rho$, i.e., when $\mathbb{E}[T]\bar{k}_{\mathrm{out}}=1$, with $\bar{k}_{\mathrm{out}}$ derived in terms of $s, \rho$, and $\beta$ in Appendix~\ref{app:gsbm-kbarout-kbar}.  In the third column, the experimentally determined critical value $\rho_c$ is compared with the corresponding theoretical prediction obtained from solving the equation $\mathbb{E}[T]\bar{k}_{\mathrm{out}}=1$ for $\rho$ after fixing $s$ and $\beta$.}
    \label{fig:geom}
\end{figure*}

A sample sequence of connection-probability maps for this SBMS is shown in Fig.~\ref{fig:geom}(a)--(c), illustrating that the connection probabilities are largest on the diagonal and decay with the cyclic distance.

To compute these block connection probabilities explicitly, set $z=\rho/s$. Averaging the geometric kernel over pairs of points in the corresponding blocks gives
\begin{equation}\label{eq:geom-explicit-block-connection-probs}
\begin{aligned}
    \bar{p}(m) &= \begin{cases}
        \frac{2\beta(2-\beta)z+\beta(\beta-1)z^2-2z^{\beta}}{(\beta-1)(2-\beta)} &\text{ if } m = 0, \\
        \frac{2(1-2^{1-\beta})z^{\beta}-\frac{1}{2}\beta(\beta-1)z^2}{(\beta-1)(2-\beta)} &\text{ if } m = 1, \\
        \frac{z^{\beta}(2m^{2-\beta}-(m-1)^{2-\beta}-(m+1)^{2-\beta})}{(\beta-1)(2-\beta)} &\text{ if } m \ge 2,
    \end{cases}
\end{aligned}
\end{equation}
as we show in Appendix~\ref{app:gsbm-pbar}. The intracommunity connection probability here is given by $\bar{p}(0)$ above, while the intercommunity connection probability between two communities at cyclic block-distance $m$ is $\bar{p}(m)$. For brevity, we henceforth write $\bar{p}(0)$ and $\bar{p}(m)$.

Vertices in the same local geometric neighborhood have a nonvanishing probability of being mutually connected. Short cycles therefore persist in the neighborhoods of vertices, so the sequence is not locally tree-like and the linearized self-consistency method does not apply. The same local structure keeps the average local clustering coefficient bounded away from zero. Figures~\ref{fig:geom}(d),~\ref{fig:geom}(g), and~\ref{fig:geom}(j) compare the clustering measured at the critical point with the estimate from Eq.~\eqref{eq:general-sbm-local-clustering-unconditioned}, while Appendix~\ref{app:clustering-derivations} demonstrates that its limiting value is strictly positive. Unlike the dense-community SBMS in \mbox{Sec.~\ref{sec:dense-community-sbm}}, the geometric model also fails to satisfy the intercommunity-sparsity assumption underlying the community-scale branching process because some intercommunity probabilities remain nonzero as $n \to \infty$. We nevertheless use the branching-process calculation as an approximation to the percolation threshold.

For this approximation, we first ignore all edges leaving a given community and fix a vertex $v$. Let $T$ denote the size of the connected component containing $v$ within its own community. Its expectation is given by Eqs.~\eqref{eq:expected-size-connected-component-linear}--\eqref{eq:prob-actually-connected-linear}, with $p$ replaced by the intracommunity connection probability $\bar{p}(0)$. Together with the value of $\bar{k}_{\mathrm{out}}$ derived in Appendix~\ref{app:gsbm-kbarout-kbar}, the branching-process condition becomes
\begin{equation}\label{eq:percolation-threshold-gsbms-general}
    \mathbb{E}[T]\bar{k}_{\mathrm{out}} = 1.
\end{equation}
This equation implicitly determines $\rho_c$ for each fixed $\beta$, with both quantities evaluated at $\rho = \rho_c$.

We next compute the average degree at the predicted threshold. A vertex has expected intracommunity degree $(s-1)\bar{p}(0)$ and expected intercommunity degree $\bar{k}_{\mathrm{out}}$, so
\begin{equation}
\begin{aligned}
    \bar{k}_c &= (s-1)\bar{p}(0) + \bar{k}_{\mathrm{out}} \\
    &= (s-1)\bar{p}(0) + \frac{1}{\mathbb{E}[T]},
\end{aligned}
\end{equation}
where both $\bar{p}(0)$ and $\mathbb{E}[T]$ are evaluated at $\rho = \rho_c$. Thus, the branching-process approximation does not generally predict a critical average degree equal to $1$.

Numerical simulations are performed at parameter values $\beta \in \{1.3, 1.5, 1.7\}$, with $\rho$ serving as the control parameter. For these three values of $\beta$, we vary $\rho$ over $[0.01, 0.41]$, $[0.1, 0.5]$, and $[0.31, 0.71]$, respectively. We fix the community size at $s=16$, and the graph size increases in powers of $2$ from $128$ to $131{,}072$. Figs.~\ref{fig:geom}(e),~\ref{fig:geom}(h), and~\ref{fig:geom}(k) show the relative size of the largest connected component, $\xi_{\mathrm{exp}}$, and the susceptibility, $\chi_{\mathrm{exp}}$, as functions of the average degree $\bar{k}$.

The branching-process predictions for $\rho_c$ are obtained by numerically solving Eq.~\eqref{eq:community-branching-threshold}, with $\bar{k}_{\mathrm{out}}$ evaluated from the expression derived in Appendix~\ref{app:gsbm-kbarout-kbar} and $\mathbb{E}[T]$ computed from the recurrence in Eqs.~\eqref{eq:expected-size-connected-component-linear}--\eqref{eq:prob-actually-connected-linear}. Figures~\ref{fig:geom}(f),~\ref{fig:geom}(i), and~\ref{fig:geom}(l) show that the accuracy of the community-scale branching-process approximation depends strongly on $\beta$. For $\beta=1.3$, where the clustering is relatively low, the experimentally determined critical value $\rho_c$ approaches the theoretical prediction closely as $n$ increases. The agreement remains reasonably good for $\beta=1.5$, although a small discrepancy persists. For $\beta=1.7$, where the clustering is substantially larger, the predicted and experimental critical points remain clearly separated even at the largest graph sizes. This loss of accuracy is consistent with the approximation underlying Eq.~\eqref{eq:percolation-threshold-gsbms-general} because as $\beta$ increases, connections become more strongly concentrated among nearby vertices, so intercommunity connections become increasingly correlated and are thus less accurately represented by a branching process between independent components. Therefore, the community-scale branching process method provides a better approximation in the lower-clustering regime, while its accuracy deteriorates as $\beta$ and the resulting clustering increase.

\section{Conclusion}\label{sec:conclusion}

In summary, we have studied the emergence of the giant connected component in five stochastic block model sequences in which the number and sizes of communities and the connection probabilities scale differently with the network size. For the three locally tree-like SBM sequences, the linearized self-consistency method predicts the percolation threshold through the condition $\lambda_{\max}=1$, in agreement with numerical simulations. For the Erdős–Rényi model with dense planted communities, where clustering remains nonzero but intercommunity connections are sparse, the threshold $\mathbb{E}[T]\bar{k}_{\mathrm{out}}=1$ obtained through the community-scale branching process method likewise agrees with simulations. For the geometric SBM, this branching process condition is only approximate, however: the agreement is strong for smaller values of $\beta$ in Eq.~\eqref{eq:geom-kernel} but deteriorates for larger values of $\beta$, where geometric correlations between communities are stronger. Together, these results show that the average degree alone does not generally determine the percolation threshold in SBM sequences. Rather, the transition depends on how edges are distributed across the community structure, and the condition $\bar{k}=1$ is recovered only in special cases.

The geometric SBM also connects the SBM sequences considered here to continuum percolation models, including random geometric graphs and soft random geometric graphs (SRGGs)~\cite{penrose1991_continuumpercolationmodel, penrose2022_Giantcomponentsoft, dubin2026_GeometryGiantComponent, penrose2016_ConnectivitySoftRandom}. The geometric SBM can be interpreted as a coarse-graining of an SRGG: geometric space is divided into intervals, and the connection probability between two communities is obtained by averaging the underlying geometric connection kernel over the corresponding pair of intervals. If the spatial width of these intervals is allowed to shrink to zero relative to the connection scale $\rho$ in Eq.~\eqref{eq:geom-kernel}, the block-averaged probabilities approach the original geometric kernel. In this sense, increasingly fine geometric SBMs provide a natural connection between the SBM framework considered here and the underlying SRGG. Recent work on random-connection and related continuum-percolation models has established sharpness of the transition or subcritical exponential decay under several broad sets of assumptions on the connection function~\cite{kupper2026_Largestcomponentsharpness, higgs2025_Exponentialdecayrandom, caicedo2025_Sharpnesspercolationphase, chebunin2026_StrongSharpPhase}. These results characterize behavior near and below the transition, but they do not identify the critical point analytically for the one-dimensional model with power-law connection probabilities considered here.

Two open problems remain. First, can analytical methods be developed that provide accurate approximations to the percolation threshold for SBM sequences with nonvanishing clustering and nonsparse intercommunity connections, a regime in which neither of the two methods considered in this paper provides a natural approximation? In particular, can such methods account for the correlations between nearby communities that are neglected by the community-scale branching process method? This could provide a more accurate description of the geometric SBM at larger values of $\beta$. Second, the geometric SBM is closely related to long-standing questions in one-dimensional long-range percolation, in both discrete and continuum formulations~\cite{newman1986_Onedimensional1, aizenman1986_Discontinuitypercolationdensity, schulman1983_Longrangepercolation, gori2017_OnedimensionalLongrangePercolation, sonmez2021_GraphDistancesContinuum}. Recent work has substantially clarified critical behavior across different regimes~\cite{hutchcroft2025_CriticalLongrangePercolation1, hutchcroft2025_CriticalLongrangePercolation2, hutchcroft2025_CriticalLongrangePercolation3, hutchcroft2021_Powerlawboundscritical}, while related locality questions have been studied through the passage from local to global giant components~\cite{alonso2026_LocalGiantsLocality}. Existing results for closely related one-dimensional power-law models provide only partial information about the location of the transition and do not yield an analytical critical curve for the SRGG studied here~\cite{schulman1983_Longrangepercolation, gori2017_OnedimensionalLongrangePercolation, coupette2025_universallyapplicableapproach}. Determining the critical value $\rho_c$ as a function of $\beta$ therefore remains an open problem.

\section{Code Availability}

The code for our simulations is available at~\url{https://github.com/moritz-laber/sbm-percolation}.

\begin{acknowledgments}

We thank \mbox{S.\,V. Scarpino}, \mbox{L. H\'ebert-Dufresne},
\mbox{F. Radicchi}, \mbox{R. van der Hofstad},
\mbox{M. Harel}, and \mbox{T. Hutchcroft} for useful discussions
and suggestions. This work was supported by NSF grant No.~CCF-2311160.

\end{acknowledgments}

\appendix

\section{SBM sampling algorithm}\label{app:algorithm}

Na\"ively sampling graphs with $n$ nodes from an edge-independent random graph model takes $\Theta(n^2)$ time as it requires sampling a Bernoulli random variable for each of the $\binom{n}{2}$ potential edges.

If the connection probabilities $p_{ij}$ are constant, i.e., $p_{ij}=p$, a faster algorithm exists~\cite{batagelj2005_efficientgeneration}. First, an arbitrary order of the potential edges is chosen, e.g., lexicographic order. Then the number of non-edges between successive realized edges is drawn from the geometric distribution with success probability $p$ until the list of potential edges is exhausted. We refer to the number of non-edges between two successive realized edges as a gap. As the number of failures before the next success in independent Bernoulli trials follows a geometric distribution, this algorithm samples from the correct distribution over graphs. The algorithm has expected time complexity $\Theta(\bar{m})$, where the expected number of edges, $\bar{m}=p n(n-1)/2$, is linear in the number of nodes $n$ for ultra-sparse graphs, $p\sim 1/n$. 

The Miller--Hagberg (MH) algorithm~\cite{miller2011_efficientgenerationnetworks} extends this idea to the Chung--Lu model~\cite{chung2002_averagedistances}, where $p_{ij}$ are not constant. A key insight from this algorithm is that if the connection probabilities $p_{ij}$ are sorted in decreasing order, then drawing gaps from the geometric distribution can be combined with rejection sampling to sample all edges in $\Theta(\bar{m})$ expected time.

We build on this central insight from the MH algorithm to sample graphs from SBMs. We provide pseudocode for this procedure in Algorithm~\ref{alg:SBM-MH} and explain key steps below.

As the connection probability is constant for a given pair of communities, the connection probabilities can be put into decreasing order by sorting $q(q+1)/2$ values. This can be done in $\mathcal{O}(q^2 \log q)$ time.

A graph $G=(\mathcal{V},\mathcal{E})$ with node set $\mathcal{V}$ and edge set $\mathcal{E}$ is then sampled by drawing gaps from the geometric distribution and using rejection sampling. After a candidate edge $e=(i,j)$ with index $\tau$ between communities $r$ and $s$ has been considered, a step $\delta \sim \mathrm{Geom}(p)$ is drawn from the geometric distribution with success probability $p=p_{rs}$. The next candidate edge $e'=(i', j')$ has index $\tau + 1+ \delta$ and would connect communities $r'$ and $s'$. As community pairs are traversed in order of decreasing probability, we know that $p_{rs}=p \geq p' = p_{r's'}$. This means the success probability $p$ used to sample $\delta$ is potentially too large. However, using rejection sampling, i.e., by accepting the candidate edge only with probability $p_{r's'}/p_{rs}$, we can correct this. Conditioned on all previous sampling outcomes $\mathcal{H}_{\tau}$ up to index $\tau$, the probability $\mathbb{P}(e' \in \mathcal{E} \mid \mathcal{H}_{\tau})$ that the edge $e'$ is realized is the product of the probability that it is selected as a candidate edge, i.e., $p_{rs}$, and the probability that the candidate edge is accepted, i.e., $p_{r's'}/p_{rs}$, and thus

\begin{equation}
    \mathbb{P}(e' \in \mathcal{E} \mid \mathcal{H}_{\tau})
    =
    p_{rs}\frac{p_{r's'}}{p_{rs}}
    =
    p_{r's'}
    .
\end{equation}

This expression does not depend on the previous sampling outcomes $\mathcal{H}_{\tau}$, and thus yields the independent Bernoulli trials for each potential edge.

Whether or not the candidate edge is accepted, the next gap is sampled with success probability $p_{r',s'}$. 

The algorithm uses two subroutines, $\mathtt{get}\_\mathtt{prob}$, which, given the connection probabilities for all community pairs and an edge index, identifies the probability associated with that edge, and $\mathtt{get}\_\mathtt{edge}$, which computes the node pair corresponding to a fixed edge index. Calls to these subroutines contribute $\mathcal{O}(\bar{m} + q^2)$ to the expected time complexity over the entire sampling procedure. The entire algorithm has expected time complexity $\mathcal{O}(\bar{m}+q^2\log q)$. Therefore, the time complexity as a function of the number of nodes $n$ depends on the graph's sparsity and the scaling of the number of communities $q$ with $n$.

\begin{algorithm}
\caption{Sampling from SBMs}\label{alg:SBM-MH}
\KwData{$\boldsymbol{n} = \{n_r\}_{r=1}^q$, $\boldsymbol{p}=\{p_{rs}\}_{1\leq r \leq s\leq q}$}
\KwResult{$G=(\mathcal{V},\mathcal{E})$}
$n \leftarrow \sum_{r=1}^q n_r$\;
$\mathcal{V} \leftarrow \{1, \dots, n\}$\;
$\mathcal{E} \leftarrow \emptyset$\;
$\boldsymbol{\tilde{p}} \leftarrow \mathtt{sort}\_\mathtt{descending}\left(\left\{ p_{rs} \right\}_{1\leq r \leq s\leq q} \right)$\;
$\tau \leftarrow 0$\;
$p \leftarrow \mathtt{get}\_\mathtt{prob}(\boldsymbol{\tilde{p}}, 1)$\;
\While{$\tau < \binom{n}{2} $}{
  \If{$p = 0$}{
    \textbf{break}\;
  }
  $\delta \sim \mathrm{Geom}(p)$\;
  $\tau \leftarrow \tau + 1 + \delta$\;
  \eIf{$\tau \leq \binom{n}{2}$}{
    $p' \leftarrow \mathtt{get}\_\mathtt{prob}(\boldsymbol{\tilde{p}},\tau)$\;
    $x \sim \mathrm{Bernoulli}\left(\frac{p'}{p}\right)$\;
        \If{$x = 1$}{
        $\mathcal{E} \leftarrow \mathcal{E}\cup \mathtt{get}\_\mathtt{edge}(\tau) $\;
    }
  }{
    \textbf{break}\;
  }
  $p \leftarrow p'$\; 
}
$G \leftarrow (\mathcal{V}, \mathcal{E})$\;
\end{algorithm}

\section{Average local clustering in the clustered SBM Sequences}
\label{app:clustering-derivations}

In this appendix, we compute the average local clustering coefficient for the Erdős--Rényi SBMS with dense planted communities and the geometric SBMS and show that it remains positive for both of them as $n \to \infty$.

\subsection{Erdős--Rényi SBMS with dense planted communities}
\label{app:clustering-dense-er-sbms}

Fix a vertex $v$ in community $r$. Let $X$ be the number of neighbors of $v$ inside community $r$, let $Y_a$ be the number of neighbors of $v$ in community $a \neq r$, and let $Y=\sum_{a \neq r} Y_a$ be the number of neighbors of $v$ outside community $r$. Then, we have $X \sim \text{Bin}(s-1, p), Y_a \sim \text{Bin}(s, b/n)$, and $Y \sim \text{Bin}((q-1)s, b/n)$, where $\text{Bin}$ denotes the binomial distribution. Since $qs=n$, this tells us that $Y$ converges in distribution to $\text{Pois}(b)$ in the $n \to \infty$ limit, where $\text{Pois}$ denotes the Poisson distribution.

Now, conditioned on the neighborhood of $v$, the expected number of edges among neighbors of $v$ is
\begin{equation}\label{eq:exp-intra-neighbor-edges}
   p\binom{X}{2} + \left(\frac{b}{n}\right)XY + p\sum_{a \neq r} \binom{Y_a}{2} + \frac{b}{n}\sum_{\underset{\scriptstyle a, a' \neq r}{\scriptstyle a<a'}} Y_a Y_{a'}.
\end{equation}
Here, the first term counts edges between two neighbors inside community $r$, the second term counts edges between one neighbor in community $r$ and one outside neighbor, the third term counts edges between two outside neighbors lying in the same outside community, and the fourth term counts edges between two outside neighbors lying in different outside communities.

For the second term in Eq.~\eqref{eq:exp-intra-neighbor-edges}, we have
\begin{equation}\label{eq:second-term-asymptotics}
\begin{aligned}
    \mathbb{E}\left[\left(\frac{b}{n}\right)XY\right] &= \frac{b}{n}\mathbb{E}[X]\mathbb{E}[Y] \\
    &= \left(\frac{b}{n}\right)\left((s-1)p\right)\left(\frac{bs(q-1)}{n}\right) \\
    &= \mathcal{O}\left(\frac{1}{n}\right).
\end{aligned}
\end{equation}
For the third term in Eq.~\eqref{eq:exp-intra-neighbor-edges}, we get
\begin{equation}\label{eq:third-term-asymptotics}
\begin{aligned}
    \mathbb{E}\left[p\sum_{a \neq r} \binom{Y_a}{2}\right] &= p(q-1)\binom{s}{2}\left(\frac{b}{n}\right)^2 \\
    &= \mathcal{O}\left(\frac{1}{n}\right).
\end{aligned}
\end{equation}
For the fourth term in Eq.~\eqref{eq:exp-intra-neighbor-edges}, we obtain
\begin{equation}\label{eq:fourth-term-asymptotics}
\begin{aligned}
    \mathbb{E}\left[\frac{b}{n}\sum_{\underset{\scriptstyle a, a' \neq r}{\scriptstyle a<a'}} Y_a Y_{a'}\right] &= \frac{b}{n} \sum_{\underset{\scriptstyle a, a' \neq r}{\scriptstyle a<a'}} \mathbb{E}[Y_a]\mathbb{E}[Y_{a'}] \\
    &= \frac{b}{n}\binom{q-1}{2} \left(\frac{bs}{n}\right)^2 \\
    &= \mathcal{O}\left(\frac{1}{n}\right).
\end{aligned}
\end{equation}
Therefore, these three terms contribute only $\mathcal{O}(n^{-1})$ to the average local clustering coefficient, so it follows that
\begin{equation}\label{eq:limiting-avg-clustering-expectation}
\begin{aligned}
    \bar{c}_n &= \mathbb{E}\left[ \frac{p\binom{X}{2}}{\binom{X+Y}{2}}  \ \middle\vert \ X+Y \ge 2 \right] + \mathcal{O}\left(\frac{1}{n}\right) \\
    &\overset{n \to \infty}{\longrightarrow} \mathbb{E}\left[ \frac{p\binom{X}{2}}{\binom{X+Y}{2}}  \ \middle\vert \ X+Y \ge 2 \right] \\
    &= \frac{\mathbb{E}\left[ \tfrac{p\binom{X}{2}}{\binom{X+Y}{2}}\mathds{1}_{\{ X+Y \ge 2\}}\right]}{\mathbb{P}(X+Y \ge 2)}.
\end{aligned}
\end{equation}
Substituting the distributions of $X$ and $Y$ and then using the fact that $\tbinom{x}{2}=0$ when $x<2$, this becomes
\begin{equation}\label{eq:limiting-avg-clustering}
\begin{aligned}
    \lim_{n \to \infty} \bar{c}_n &= \frac{\sum_{x=0}^{s-1}\sum_{y=0}^{\infty} \binom{s-1}{x}p^x (1-p)^{s-1-x} e^{-b} \frac{b^y}{y!} \frac{p\binom{x}{2}}{\binom{x+y}{2}}\mathds{1}_{x+y \ge 2}}{\mathbb{P}(X+Y \ge 2)} \\
    &= \frac{p\sum_{x=2}^{s-1} \sum_{y=0}^{\infty} \binom{s-1}{x} p^x (1-p)^{s-1-x} e^{-b}\frac{b^y}{y!} \frac{x(x-1)}{(x+y)(x+y-1)}}{\mathbb{P}(X+Y \ge 2)}.
\end{aligned}
\end{equation}
It remains to compute $\mathbb{P}(X+Y \ge 2)$. Note that
\begin{equation}\label{eq:complementary-probability}
    \mathbb{P}(X+Y \ge 2) = 1 - \mathbb{P}(X+Y=0) - \mathbb{P}(X+Y=1)
\end{equation}
Then, we have $\mathbb{P}(X+Y=0) = e^{-b}(1-p)^{s-1}$, and considering separately the cases $X=1$ or $Y=1$, we get
\begin{equation}\label{eq:prob-one-neighbor}
    \mathbb{P}(X+Y=1) = e^{-b}(s-1)p(1-p)^{s-2} + be^{-b} (1-p)^{s-1}.
\end{equation}
Thus, substituting the resulting expression for $\mathbb{P}(X+Y \ge 2)$ into Eq.~\eqref{eq:limiting-avg-clustering} and using the fact that $1/\tbinom{d}{2} = 2\int_0^1 (1-u)u^{d-2}  \ du$ for all $d \ge 2$ yields
\begin{widetext}
\begin{equation}\label{eq:limiting-avg-clustering-final}
\begin{aligned}
    \lim_{n \to \infty} \bar{c}_n &= \frac{p\sum_{x=2}^{s-1} \sum_{y=0}^{\infty} \binom{s-1}{x} p^x (1-p)^{s-1-x} e^{-b}\frac{b^y}{y!} \frac{x(x-1)}{(x+y)(x+y-1)}}{1 - e^{-b}(1-p)^{s-2}(1-p + (s-1)p + b(1-p))} \\
    &= \frac{(s-1)(s-2)p^3\int_0^1 (1-u)(1-p+pu)^{s-3} e^{-b(1-u)} \ du}{1 - e^{-b}(1-p)^{s-2}(1-p + (s-1)p + b(1-p))} \\
    &\ge \frac{pe^{-b}\tbinom{s-1}{2}p^2(1-p)^{s-3}}{1 - e^{-b}(1-p)^{s-2}(1-p + (s-1)p + b(1-p))} \\
    &> 0,
\end{aligned}
\end{equation}
\end{widetext}
where the last bound is taken by considering only the event that $v$ has exactly two neighbors inside community $r$ and no outside neighbors. Therefore, the Erdős--Rényi SBMS with dense planted communities indeed has nonvanishing average local clustering.

As a check for the clustering expression, note that if $s=1$ or $s=2$, then $\lim_{n \to \infty} \bar{c}_n = 0$, as expected, and if $b=0$, then $\lim_{n \to \infty} \bar{c}_n = p$, which is also as expected.

\subsection{Geometric SBMS}
\label{app:clustering-gsbms}

Let $X_a$ be the number of neighbors of $v$ in community $a$. Then, we know that $X_r \sim \text{Bin}(s-1, \bar{p}(0))$, and for $a \neq r$, we have $X_a \sim \text{Bin}(s, \bar{p}(\min\{|a-r|, q-|a-r|\}))$, where $\text{Bin}$ denotes the binomial distribution and $\min\{|a-r|, q-|a-r|\}$ is the shortest cyclic distance between communities $a$ and $r$.

Now, conditioned on the values of $X_1, X_2, \ldots, X_q$, the expected number of edges among the neighbors of $v$ is
\begin{equation}\label{eq:exp-intra-neighbor-edges-gsbm}
    \bar{p}(0)\sum_{a=1}^q \binom{X_a}{2} + \sum_{1 \le a < a' \le q} \bar{p}(\min\{|a-a'|, q-|a-a'|\}) X_a X_{a'}.
\end{equation}
Also, as $n \to \infty$, we have $q \to \infty$, so any fixed neighborhood of community $r$ is eventually unaffected by the cyclic wrap-around because the contribution from communities whose distance grows with $q$ is negligible due to the fact that $\sum_{m \ge 1} \bar{p}(m) < \infty$. Hence, the limiting average local clustering coefficient is
\begin{widetext}
\begin{equation}\label{eq:avg-clustering-gsbm-finite}
\begin{aligned}
    \bar{c}_n &= \mathbb{E}\left[\frac{\bar{p}(0)\sum_{a=1}^q \binom{X_a}{2} + \sum_{1 \le a < a' \le q} \bar{p}(\min\{|a-a'|, q-|a-a'|\}) X_a X_{a'}}{\tbinom{\sum_{a=1}^q X_a}{2}} \ \middle\vert \ \sum_{a=1}^q X_a \ge 2\right] \\
    &\overset{n \to \infty}{\longrightarrow} \mathbb{E}\left[ \frac{\bar{p}(0)\sum_{a \in \mathbb{Z}}\tbinom{X_a}{2}+\sum_{a<a'} \bar{p}(|a-a'|)X_aX_{a'}}{\tbinom{\sum_{a \in \mathbb{Z}}X_a}{2}} \ \middle\vert \ \sum_{a \in \mathbb{Z}}X_a \ge 2\right],
\end{aligned}
\end{equation}
\end{widetext}
where $X_r \sim \text{Bin}(s-1, \bar{p}(0))$, and for $a \neq r$, $X_a \sim \text{Bin}(s, \bar{p}(|a-r|))$.

Unlike in the Erdős--Rényi SBMS with dense planted communities, the terms involving outside communities cannot be discarded here. In the Erdős--Rényi SBMS, any edge involving an outside-community neighbor has probability $\mathcal{O}(n^{-1})$, so all such triangle-closing contributions vanish asymptotically. In the geometric SBMS, however, two outside-community neighbors can lie in nearby communities $a$ and $a'$, and the probability that they are connected is $\bar{p}(|a-a'|)$, which is independent of $n$.

It remains to show that the limiting clustering coefficient is bounded away from $0$. Consider the event that $v$ has exactly two neighbors inside community $r$ and no neighbors outside community $r$. In the limiting setting, this event has probability
\begin{equation}\label{eq:prob-two-inside-neighbors}
    \binom{s-1}{2} \bar{p}(0)^2 (1-\bar{p}(0))^{s-3}\prod_{m=1}^{\infty} (1-\bar{p}(m))^{2s}.
\end{equation}
On this event, $v$ has degree exactly $2$, and both neighbors lie in community $r$. These two neighbors are connected with probability $\bar{p}(0)$. Thus, since the conditioning event $\sum_{a \in \mathbb{Z}} X_a \ge 2$ has probability at most $1$, we obtain
\begin{equation}
    \lim_{n \to \infty} \bar{c}_n \ge  \binom{s-1}{2} \bar{p}(0)^3 (1-\bar{p}(0))^{s-3}\prod_{m=1}^{\infty} (1-\bar{p}(m))^{2s}.
\end{equation}
Now, for large $m$, we have
\begin{equation}
    \bar{p}(m) = \left(\frac{\rho}{s}\right)^{\beta} m^{-\beta} + \mathcal{O}(m^{-\beta-2}).
\end{equation}
(See Appendix~\ref{app:gsbm-pbar} for the full calculations for $\bar{p}(m)$.) Since $\beta>1$, we have $\sum_{m=1}^{\infty} \bar{p}(m) < \infty$, so it follows that $\prod_{m=1}^{\infty} (1-\bar{p}(m))^{2s}$ converges to a positive constant. Also, $\bar{p}(0)>0$. Hence, we can conclude that $\lim_{n \to \infty} \bar{c}_n > 0$, so the geometric SBMS has nonvanishing average local clustering in the thermodynamic limit.

\section{Block-averaged connection probabilities in the geometric SBMS}
\label{app:gsbm-pbar}

In this appendix, we derive the block-averaged connection probabilities from Eq.~\eqref{eq:geom-explicit-block-connection-probs} used in the geometric SBM sequence. Throughout this calculation, we write
\begin{equation}\label{eq:z-def}
    z = \frac{\rho}{s}.
\end{equation}
Since we imposed the restriction $\rho < s$, observe that $0 < z < 1$.

We first compute the intracommunity connection probability $\bar{p}(0)$. This is the average of $p(|x-y|)$ over two points $x, y$ chosen uniformly from the same interval of length $s$, giving
\begin{equation}\label{eq:pbar0-double-integral}
    \bar{p}(0) = \frac{1}{s^2} \int_0^s \int_0^s p(|x-y|) \ dy \ dx.
\end{equation}
By symmetry across the diagonal $x=y$, we can rewrite this to remove the absolute value signs, giving
\begin{equation}\label{eq:pbar-symmetry}
    \bar{p}(0) = \frac{2}{s^2} \int_0^s \int_0^x p(x-y) \ dy \ dx.
\end{equation}
Setting $r=x-y$ and then changing the order of integration then yields
\begin{equation}\label{eq:pbar0-distance-integral}
\begin{aligned}
    \bar{p}(0) &= \frac{2}{s^2}\int_0^s \int_0^x p(r) \ dr \ dx \\
    &= \frac{2}{s^2} \int_0^s (s-r) p(r) \ dr.
\end{aligned}
\end{equation}
Now, set $r=st$. Since $p(st) = \min\{1, (z/t)^{\beta}\}$, we get
\begin{equation}\label{eq:pbar0-scaled-integral}
    \bar{p}(0) = 2\int_0^1 (1-t)\min\left\{ 1, \left(\frac{z}{t}\right)^{\beta} \right\} \ dt.
\end{equation}
Splitting the integral at $t=z$ (using the fact that $0 < z < 1$), this becomes
\begin{equation}\label{eq:pbar0-split}
    \bar{p}(0) = 2\int_0^z (1-t) \ dt + 2z^{\beta} \int_z^1 (1-t)t^{-\beta} \ dt.
\end{equation}
Evaluating each of these two terms using standard methods yields
\begin{equation}\label{eq:pbar0-integral-values}
\begin{aligned}
    2\int_0^z (1-t) \ dt &= 2z - z^2, \\
    2z^{\beta}\int_z^1 (1-t)t^{-\beta} \ dt &= \frac{2(\beta-1)z^2-2(\beta-2)z - 2z^{\beta}}{(\beta-1)(2-\beta)}.
\end{aligned}
\end{equation}
Therefore, it follows that the intracommunity connection probability is given by
\begin{equation}\label{eq:pbar0-final}
\begin{aligned}
    \bar{p}(0) &= 2z - z^2 + \frac{2(\beta-1)z^2-2(\beta-2)z - 2z^{\beta}}{(\beta-1)(2-\beta)} \\
    &= \frac{2\beta(2-\beta)z + \beta(\beta-1)z^2-2z^{\beta}}{(\beta-1)(2-\beta)}.
\end{aligned}
\end{equation}
Next, we compute the connection probability $\bar{p}(1)$ between two adjacent communities. Taking the first community to be $[0, s]$ and the second to be $[s, 2s]$, we have
\begin{equation}\label{eq:pbar1-double-integral}
    \bar{p}(1) = \frac{1}{s^2} \int_0^s \int_s^{2s} p(y-x) \ dy \ dx.
\end{equation}
Letting $r=y-x$, the possible distances range from $0$ to $2s$, and the number of pairs with distance $r$ is proportional to $r$ for $0 \le r \le s$ and $2s-r$ for $s \le r \le 2s$. Hence,
\begin{equation}\label{eq:pbar1-distance-integral}
    \bar{p}(1) = \frac{1}{s^2}\left( \int_0^s rp(r) \ dr + \int_s^{2s} (2s-r) p(r) \ dr\right).
\end{equation}
Once again, we set $r=st$, giving
\begin{equation}\label{eq:pbar1-scaled-integral}
    \bar{p}(1) = \int_0^1 tp(st) \ dt + \int_1^2 (2-t) p(st) \ dt.
\end{equation}
Since $z<1$, the first integral must split at $t=z$, while the second integral is entirely in the regime $p(st)=(z/t)^{\beta}$. Therefore, we obtain
\begin{equation}\label{eq:pbar1-split}
    \bar{p}(1) = \int_0^z t \ dt + z^{\beta} \int_z^1 t^{1-\beta} \ dt + z^{\beta} \int_1^2 (2-t)t^{-\beta} \ dt.
\end{equation}
Evaluating each of these three terms yields
\begin{equation}\label{eq:pbar1-integral-values}
\begin{aligned}
    \int_0^z t \ dt &= \frac{z^2}{2}, \\
    z^{\beta} \int_z^1 t^{1-\beta} \ dt &= \frac{z^{\beta}-z^2}{2-\beta}, \\
    z^{\beta}\int_1^2 (2-t)t^{-\beta} \ dt &= \frac{2(1-2^{1-\beta})z^{\beta}}{\beta-1} - \frac{(2^{2-\beta}-1)z^{\beta}}{2-\beta}.
\end{aligned}
\end{equation}
Adding these terms and simplifying yields
\begin{equation}\label{eq:pbar1-final}
    \bar{p}(1) = \frac{2(1-2^{1-\beta})z^{\beta}-\tfrac{1}{2}\beta(\beta-1)z^2}{(\beta-1)(2-\beta)}.
\end{equation}
Finally, we compute $\bar{p}(m)$ for $m \ge 2$. In this case, the two communities are separated far enough that every distance between them is larger than $\rho$, so the kernel is always in its power-law regime. Taking the first community to be $[0, s]$ and the second to be $[ms, (m+1)s]$, we have
\begin{equation}\label{eq:pbarm-double-integral}
    \bar{p}(m) = \frac{1}{s^2} \int_0^s \int_{ms}^{(m+1)s} p(y-x) \ dy \ dx.
\end{equation}
Again setting $r=y-x$, the possible distances range from $(m-1)s$ to $(m+1)s$. The number of pairs with distance $r$ grows linearly from $(m-1)s$ to $ms$, and then decreases linearly from $ms$ to $(m+1)s$. Therefore, splitting the double integral for $\bar{p}(m)$ into an integral from $(m-1)s$ to $ms$ and an integral from $ms$ to $(m+1)s$, using the fact that $p(r) = (\rho/r)^{\beta}$ because $r>\rho$ holds throughout the range of integration, and setting $r=st$ as before, we get
\begin{widetext}
\begin{equation}\label{eq:pbarm-final}
\begin{aligned}
    \bar{p}(m) &= \frac{1}{s^2}\left(\int_{m-1)s}^{ms} (r-(m-1)s)p(r) \ dr + \int_{ms}^{(m+1)s} ((m+1)s-r)p(r) \ dr\right) \\
    &= z^{\beta} \int_{m-1}^m (t-m+1)t^{-\beta} \ dt + z^{\beta} \int_m^{m+1} (m+1-t)t^{-\beta} \ dt \\
    &= z^{\beta}\left(\frac{m^{2-\beta}-(m-1)^{2-\beta}}{2-\beta} + \frac{(m-1)m^{1-\beta}-(m-1)^{2-\beta}}{\beta-1}\right) \\
    &\quad + z^{\beta}\left(\frac{(m+1)m^{1-\beta} - (m+1)^{2-\beta}}{\beta-1} - \frac{(m+1)^{2-\beta}-m^{2-\beta}}{2-\beta}\right) \\
    &= \frac{z^{\beta}(2m^{2-\beta}-(m-1)^{2-\beta}-(m+1)^{2-\beta})}{(\beta-1)(2-\beta)}.
\end{aligned}
\end{equation}
\end{widetext}

Combining Eq.~\eqref{eq:pbar0-final}, Eq.~\eqref{eq:pbar1-final}, and Eq.~\eqref{eq:pbarm-final} gives Eq.~\eqref{eq:geom-explicit-block-connection-probs}.

\section{Average out-degree and average degree in the geometric SBMS}
\label{app:gsbm-kbarout-kbar}

We now derive the average out-degree $\bar{k}_{\mathrm{out}}$, meaning the expected number of neighbors of a vertex outside of its own community, and the total average degree $\bar{k}$ for the geometric SBMS.

Fix a vertex in some community. For each cyclic block-distance $m \ge 1$, there are two communities at distance $m$, one in each direction around the circle, and each one contains $s$ vertices. Since the connection probability to each vertex in either community is $\bar{p}(m)$, the expected number of neighbors in communities at distance $m$ is $2s\bar{p}(m)$. Therefore, in the thermodynamic limit,
\begin{equation}\label{eq:average-out-deg}
    \bar{k}_{\mathrm{out}} = 2s\sum_{m=1}^{\infty} \bar{p}(m).
\end{equation}
Using Eq.~\eqref{eq:geom-explicit-block-connection-probs}, the first term in this sum is
\begin{equation}
    \bar{p}(1) = \frac{2(1-2^{1-\beta})z^{\beta}-\tfrac{1}{2}\beta(\beta-1) z^2}{(\beta-1)(2-\beta)},
\end{equation}
while, for $m \ge 2$,
\begin{equation}
    \bar{p}(m) = \frac{z^{\beta}(2m^{2-\beta}-(m-1)^{2-\beta}-(m+1)^{2-\beta})}{(\beta-1)(2-\beta)},
\end{equation}
where in both of these expressions, $z = \rho/s$. Hence, we obtain
\begin{widetext}
\begin{equation}\label{eq:summation-pbar}
\begin{aligned}
    \sum_{m=1}^{\infty} \bar{p}(m) &= \bar{p}(1) + \sum_{m=2}^{\infty} \bar{p}(m) \\
    &= \frac{2(1-2^{1-\beta})z^{\beta}-\tfrac{1}{2}\beta(\beta-1) z^2}{(\beta-1)(2-\beta)} \\
    &\quad +  \frac{z^{\beta}}{(\beta-1)(2-\beta)} \sum_{m=2}^{\infty} (2m^{2-\beta} - (m-1)^{2-\beta} - (m+1)^{2-\beta}).
\end{aligned}
\end{equation}
\end{widetext}
Let $a_m = m^{2-\beta}$ for all $m \ge 1$. Then, the summand in the last expression above is equal to $2a_m - a_{m-1} - a_{m+1}$, which can be rewritten as $(a_m-a_{m-1})-(a_{m+1}-a_m)$. Thus, the sum telescopes, i.e.,
\begin{equation}
\begin{aligned}
    \sum_{m=2}^{\infty} ((a_m-a_{m-1})-(a_{m+1}-a_m))
    &= a_2 - a_1 \\
    &= 2^{2-\beta}-1.
\end{aligned}
\end{equation}
Substituting this value for the infinite summation into Eq.~\eqref{eq:summation-pbar} gives
\begin{equation}
\begin{aligned}
    \sum_{m=1}^{\infty} \bar{p}(m) &= \frac{2(1-2^{1-\beta})z^{\beta}-\tfrac{1}{2}\beta(\beta-1)z^2 + (2^{2-\beta}-1)z^{\beta}}{(\beta-1)(2-\beta)} \\
    &= \frac{z^{\beta}-\tfrac{1}{2}\beta(\beta-1)z^2}{(\beta-1)(2-\beta)}.
\end{aligned}
\end{equation}
It then follows from Eq.~\eqref{eq:average-out-deg} that
\begin{equation}\label{eq:average-out-deg-final}
    \bar{k}_{\mathrm{out}} = \frac{2sz^{\beta}-\beta(\beta-1)sz^2}{(\beta-1)(2-\beta)}.
\end{equation}
We next compute the total average degree. A vertex has $s-1$ possible neighbors inside its own community, each connected to it with probability $\bar{p}(0)$. Its expected intracommunity degree is therefore $(s-1)\bar{p}(0)$. Adding the expected degree outside its community gives
\begin{equation}\label{eq:average-deg-expression}
    \bar{k} = (s-1)\bar{p}(0) + \bar{k}_{\mathrm{out}}.
\end{equation}
From Eq.~\eqref{eq:geom-explicit-block-connection-probs}, the intracommunity connection probability is
\begin{equation}\label{eq:intracommunity-pbar-geom}
    \bar{p}(0) = \frac{2\beta(2-\beta)z + \beta(\beta-1) z^2 - 2z^{\beta}}{(\beta-1)(2-\beta)}.
\end{equation}
Substituting Eqs.~\eqref{eq:average-out-deg-final} and \eqref{eq:intracommunity-pbar-geom} into Eq.~\eqref{eq:average-deg-expression}, we obtain
\begin{equation}\label{eq:average-deg-final}
\begin{aligned}
    \bar{k} &= (s-1) \frac{2\beta(2-\beta)z+\beta(\beta-1)z^2-2z^{\beta}}{(\beta-1)(2-\beta)} \\
    &\quad + \frac{2sz^{\beta} - \beta(\beta-1)sz^2}{(\beta-1)(2-\beta)} \\
    &= \frac{2\beta(2-\beta)(s-1)z-\beta(\beta-1)z^2 + 2z^{\beta}}{(\beta-1)(2-\beta)}.
\end{aligned}
\end{equation}
Equations \eqref{eq:average-out-deg-final} and \eqref{eq:average-deg-final} give the limiting average out-degree and total average degree used in the geometric-SBMS percolation calculation.

\bibliography{references}

@article{allard2009_HeterogeneousBondPercolation,
  title = {Heterogeneous Bond Percolation on Multitype Networks with an Application to Epidemic Dynamics},
  author = {Allard, Antoine and No{\"e}l, Pierre-Andr{\'e} and Dub{\'e}, Louis J. and Pourbohloul, Babak},
  year = 2009,
  journal = {Physical Review E},
  volume = {79},
  number = {3},
  pages = {036113},
  doi = {10.1103/PhysRevE.79.036113},
  url = {https://journals.aps.org/pre/abstract/10.1103/PhysRevE.79.036113}
}

@article{allard2019_PercolationEffectiveStructure,
  title = {Percolation and the {{Effective Structure}} of {{Complex Networks}}},
  author = {Allard, Antoine and {H{\'e}bert-Dufresne}, Laurent},
  year = 2019,
  journal = {Physical Review X},
  volume = {9},
  number = {1},
  pages = {011023},
  publisher = {American Physical Society},
  doi = {10.1103/PhysRevX.9.011023},
  url = {https://link.aps.org/doi/10.1103/PhysRevX.9.011023}
}

@article{ball2008_Thresholdbehaviourfinal,
  title = {Threshold Behaviour and Final Outcome of an Epidemic on a Random Network with Household Structure},
  author = {Ball, Frank and Sirl, David and Trapman, Pieter},
  year = 2009,
  journal = {Advances in Applied Probability},
  volume = {41},
  number = {3},
  pages = {765--796},
  doi = {10.1239/aap/1253281063}
}

@article{batagelj2005_efficientgeneration,
  title = {Efficient Generation of Large Random Networks},
  author = {Batagelj, Vladimir and Brandes, Ulrik},
  year = 2005,
  month = mar,
  journal = {Physical Review E},
  volume = {71},
  number = {3},
  pages = {036113},
  publisher = {American Physical Society},
  doi = {10.1103/PhysRevE.71.036113}
}

@article{bollobas2005_phasetransitioninhomogeneous,
  title = {The Phase Transition in Inhomogeneous Random Graphs},
  author = {Bollob{\'a}s, B{\'e}la and Janson, Svante and Riordan, Oliver},
  year = 2007,
  journal = {Random Structures \& Algorithms},
  volume = {31},
  pages = {3--122},
  doi = {10.1002/rsa.20168},
  url = {https://arxiv.org/abs/math/0504589v3}
}

@article{cantwell2019_MessagePassingLoops,
  title = {Message Passing on Networks with Loops},
  author = {Cantwell, George T. and Newman, M. E. J.},
  year = 2019,
  journal = {Proceedings of the National Academy of Sciences},
  volume = {116},
  number = {47},
  pages = {23398--23403},
  doi = {10.1073/pnas.1914893116},
  url = {https://www.pnas.org/doi/10.1073/pnas.1914893116}
}

@article{chung2002_averagedistances,
  title = {The Average Distances in Random Graphs with given Expected Degrees},
  author = {Chung, Fan and Lu, Linyuan},
  year = 2002,
  month = dec,
  journal = {Proceedings of the National Academy of Sciences},
  volume = {99},
  number = {25},
  pages = {15879--15882},
  publisher = {Proceedings of the National Academy of Sciences},
  doi = {10.1073/pnas.252631999}
}

@article{cirigliano2024_ScalingUniversalityPercolation,
  title = {Scaling and Universality for Percolation in Random Networks: A Unified View},
  author = {Cirigliano, Lorenzo and Tim{\'a}r, G{\'a}bor and Castellano, Claudio},
  year = 2024,
  journal = {Physical Review E},
  volume = {110},
  number = {6},
  pages = {064303},
  doi = {10.1103/PhysRevE.110.064303},
  url = {https://arxiv.org/abs/2408.05125}
}

@article{decelle2011_AsymptoticAnalysisSBM,
  title = {Asymptotic Analysis of the Stochastic Block Model for Modular Networks and Its Algorithmic Applications},
  author = {Decelle, Aur{\'e}lien and Krzakala, Florent and Moore, Cristopher and Zdeborov{\'a}, Lenka},
  year = 2011,
  journal = {Physical Review E},
  volume = {84},
  number = {6},
  pages = {066106},
  doi = {10.1103/PhysRevE.84.066106},
  url = {https://journals.aps.org/pre/abstract/10.1103/PhysRevE.84.066106}
}

@article{do2023_ComponentBehaviourExcess,
  title = {Component {{Behaviour}} and {{Excess}} of {{Random Bipartite Graphs Near}} the {{Critical Point}}},
  author = {Do, Tuan and Erde, Joshua and Kang, Mihyun and Missethan, Michael},
  year = 2023,
  journal = {The Electronic Journal of Combinatorics},
  volume = {30},
  number = {3},
  pages = {P3.7},
  doi = {10.37236/11065},
  url = {https://www.combinatorics.org/ojs/index.php/eljc/article/view/v30i3p7}
}

@article{erdos1960_evolutionrandomgraphs,
  title = {On the Evolution of Random Graphs},
  author = {Erd{\H o}s, Paul and R{\'e}nyi, Alfr{\'e}d},
  year = 1960,
  journal = {Publications of the Mathematical Institute of the Hungarian Academy of Sciences},
  volume = {5},
  number = {1},
  pages = {17--60},
}

@article{gilbert1959_RandomGraphs,
  title = {Random {{Graphs}}},
  author = {Gilbert, E. N.},
  year = 1959,
  month = dec,
  journal = {The Annals of Mathematical Statistics},
  volume = {30},
  number = {4},
  pages = {1141--1144},
  publisher = {Institute of Mathematical Statistics},
  doi = {10.1214/aoms/1177706098},
  url = {https://projecteuclid.org/journals/annals-of-mathematical-statistics/volume-30/issue-4/Random-Graphs/10.1214/aoms/1177706098.full}
}

@article{gleeson2009_Howclusteringaffects,
  title = {How Clustering Affects the Bond Percolation Threshold in Complex Networks},
  author = {Gleeson, James P. and Melnik, Sergey and Hackett, Adam},
  year = 2010,
  month = jun,
  journal = {Physical Review E},
  volume = {81},
  pages = {066114},
  doi = {10.1103/PhysRevE.81.066114},
  url = {https://arxiv.org/abs/0912.4204v2}
}

@article{holland1983_StochasticBlockmodels,
  title = {Stochastic Blockmodels: First Steps},
  author = {Holland, Paul W. and Laskey, Kathryn Blackmond and Leinhardt, Samuel},
  year = 1983,
  journal = {Social Networks},
  volume = {5},
  number = {2},
  pages = {109--137},
  doi = {10.1016/0378-8733(83)90021-7}
}

@mastersthesis{johansson2012_giantcomponentrandom,
  title = {The giant component of the random bipartite graph},
  author = {Johansson, Tony},
  year = 2012,
  address = {G\"oteborg, Sweden},
  url = {https://publications.lib.chalmers.se/records/fulltext/170011/170011.pdf},
  school = {Chalmers University of Technology}
}

@article{kang2015_phasetransitionmultitype,
  title = {{{The Phase Transition}} in {{Multitype Binomial Random Graphs}}},
  author = {Kang, Mihyun and Koch, Christoph and Pach{\'o}n, Ang{\'e}lica},
  year = 2015,
  journal = {SIAM Journal on Discrete Mathematics},
  volume = {29},
  number = {2},
  pages = {1042--1064},
  publisher = {Society for Industrial and Applied Mathematics},
  doi = {10.1137/140973256},
  url = {https://doi.org/10.1137/140973256}
}

@article{karrer2010_Randomgraphscontaining,
  title = {Random Graphs Containing Arbitrary Distributions of Subgraphs},
  author = {Karrer, Brian and Newman, M. E. J.},
  year = 2010,
  month = dec,
  journal = {Physical Review E},
  volume = {82},
  pages = {066118},
  doi = {10.1103/PhysRevE.82.066118},
  url = {https://arxiv.org/abs/1005.1659v1}
}

@article{karrer2011_StochasticBlockmodelsCommunity,
  title = {Stochastic Blockmodels and Community Structure in Networks},
  author = {Karrer, Brian and Newman, M. E. J.},
  year = 2011,
  journal = {Physical Review E},
  volume = {83},
  number = {1},
  pages = {016107},
  doi = {10.1103/PhysRevE.83.016107},
  url = {http://dx.doi.org/10.1103/PhysRevE.83.016107}
}

@article{karrer2014_Percolationsparsenetworks,
  title = {Percolation on {{Sparse}} {{Networks}}},
  author = {Karrer, Brian and Newman, M. E. J. and Zdeborov{\'a}, Lenka},
  year = 2014,
  journal = {Physical Review Letters},
  volume = {113},
  pages = {208702},
  doi = {10.1103/PhysRevLett.113.208702},
  url = {https://arxiv.org/abs/1405.0483v2}
}

@article{li2021_PercolationComplexNetworks,
  title = {Percolation on Complex Networks: Theory and Application},
  author = {Li, Ming and Liu, Run-Ran and L{\"u}, Linyuan and Hu, Mao-Bin and Xu, Shuqi and Zhang, Yi-Cheng},
  year = 2021,
  journal = {Physics Reports},
  volume = {907},
  pages = {1--68},
  doi = {10.1016/j.physrep.2020.12.003},
  url = {https://doi.org/10.1016/j.physrep.2020.12.003}
}

@book{meurant2025_HessenbergTridiagonalMatrices,
  title = {Hessenberg and Tridiagonal Matrices: Theory and Examples},
  author = {Meurant, G{\'e}rard},
  year = {2025},
  publisher = {Society for Industrial and Applied Mathematics},
  isbn = {978-1-61197-844-5},
  doi = {10.1137/1.9781611978452},
  url = {https://epubs.siam.org/doi/book/10.1137/1.9781611978452}
}

@inproceedings{miller2011_efficientgenerationnetworks,
  title = {Efficient {{Generation}} of {{Networks}} with {{Given Expected Degrees}}},
  booktitle = {Algorithms and {{Models}} for the {{Web Graph}}},
  author = {Miller, Joel C. and Hagberg, Aric},
  editor = {Frieze, Alan and Horn, Paul and Pra{\l}at, Pawe{\l}},
  year = 2011,
  pages = {115--126},
  publisher = {Springer},
  address = {Berlin, Heidelberg},
  doi = {10.1007/978-3-642-21286-4_10},
  isbn = {978-3-642-21286-4}
}

@article{newman2000_Randomgraphsarbitrary,
  title = {Random Graphs with Arbitrary Degree Distributions and Their Applications},
  author = {Newman, M. E. J. and Strogatz, S. H. and Watts, D. J.},
  year = 2001,
  journal = {Physical Review E},
  volume = {64},
  number = {2},
  pages = {026118},
  doi = {10.1103/PhysRevE.64.026118}
}

@article{newman2000_efficientmontecarlo,
  title = {Efficient {{Monte Carlo Algorithm}} and {{High-Precision Results}} for {{Percolation}}},
  author = {Newman, M. E. J. and Ziff, R. M.},
  year = 2000,
  month = nov,
  journal = {Physical Review Letters},
  volume = {85},
  number = {19},
  pages = {4104--4107},
  publisher = {American Physical Society},
  doi = {10.1103/PhysRevLett.85.4104}
}

@article{newman2001_fastmontecarlo,
  title = {Fast {{Monte Carlo}} Algorithm for Site or Bond Percolation},
  author = {Newman, M. E. J. and Ziff, R. M.},
  year = 2001,
  month = jun,
  journal = {Physical Review E},
  volume = {64},
  number = {1},
  pages = {016706},
  publisher = {American Physical Society},
  doi = {10.1103/PhysRevE.64.016706}
}

@article{radicchi2016_BeyondLocallyTreelike,
  title = {Beyond the Locally Treelike Approximation for Percolation on Real Networks},
  author = {Radicchi, Filippo and Castellano, Claudio},
  year = 2016,
  journal = {Physical Review E},
  volume = {93},
  number = {3},
  pages = {030302},
  doi = {10.1103/PhysRevE.93.030302}
}

@article{soderberg2002_Generalformalisminhomogeneous,
  title = {General Formalism for Inhomogeneous Random Graphs},
  author = {S{\"o}derberg, Bo},
  year = 2002,
  month = dec,
  journal = {Physical Review E},
  volume = {66},
  number = {6},
  pages = {066121},
  publisher = {American Physical Society},
  doi = {10.1103/PhysRevE.66.066121},
  url = {https://link.aps.org/doi/10.1103/PhysRevE.66.066121}
}

@article{stegehuis2016_PowerLawRelationsCommunities,
  title = {Power-Law Relations in Random Networks with Communities},
  author = {Stegehuis, Clara and van der Hofstad, Remco and van Leeuwaarden, Johan S. H.},
  year = 2016,
  journal = {Physical Review E},
  volume = {94},
  number = {1},
  pages = {012302},
  doi = {10.1103/PhysRevE.94.012302},
  url = {https://arxiv.org/abs/1603.09711}
}

@article{vanderhofstad2016_HierarchicalConfigurationModel,
  title = {{{Hierarchical Configuration Model}}},
  author = {van der Hofstad, Remco and van Leeuwaarden, Johan S. H. and Stegehuis, Clara},
  year = {2016},
  month = dec,
  journal = {Internet Mathematics},
  pages = {},
  note = {Article 1214},
  doi = {10.24166/im.01.2017}
}

@misc{alonso2026_LocalGiantsLocality,
  title = {From Local Giants to Locality in Long-Range Percolation},
  author = {Moreno Alonso, Yago and Komj{\'a}thy, J{\'u}lia},
  year = 2026,
  month = jul,
  eprint = {2607.18011},
  primaryclass = {math.PR},
  publisher = {arXiv},
  doi = {10.48550/arXiv.2607.18011},
  archiveprefix = {arXiv}
}

@misc{dubin2026_GeometryGiantComponent,
  title = {The Geometry of the Giant Component of Random Geometric Graphs},
  author = {Dubin, Karoline and Gorski, Christian and Michelen, Marcus},
  year = 2026,
  month = jun,
  eprint = {2606.01627},
  primaryclass = {math.PR},
  publisher = {arXiv},
  doi = {10.48550/arXiv.2606.01627},
  archiveprefix = {arXiv}
}

@misc{franchi2026_ComponentStructurePercolation,
  title = {Component Structure and Percolation in Block Models},
  author = {Franchi, Riccardo and Newman, M. E. J.},
  year = 2026,
  month = jul,
  eprint = {2607.20719},
  primaryclass = {cs.SI},
  publisher = {arXiv},
  doi = {10.48550/arXiv.2607.20719},
  archiveprefix = {arXiv}
}

@article{bujok2014_PercolationClassicalBlockmodel,
  title = {Percolation in the Classical Blockmodel},
  author = {Bujok, Maksymilian and Fronczak, Piotr and Fronczak, Agata},
  year = 2014,
  month = sep,
  journal = {The European Physical Journal B},
  volume = {87},
  number = {9},
  pages = {212},
  doi = {10.1140/epjb/e2014-50242-x}
}

@article{allard2015_GeneralExactApproach,
  title = {General and Exact Approach to Percolation on Random Graphs},
  author = {Allard, Antoine and {H{\'e}bert-Dufresne}, Laurent and Young, Jean-Gabriel and Dub{\'e}, Louis J.},
  year = 2015,
  month = dec,
  journal = {Physical Review E},
  volume = {92},
  number = {6},
  pages = {062807},
  publisher = {American Physical Society},
  doi = {10.1103/PhysRevE.92.062807}
}

@article{vanderhofstad2022_PhaseTransitionRandom,
  title = {Phase Transition in Random Intersection Graphs with Communities},
  author = {{van der Hofstad}, Remco and Komj{\'a}thy, J{\'u}lia and Vadon, Vikt{\'o}ria},
  year = 2022,
  journal = {Random Structures \& Algorithms},
  volume = {60},
  number = {3},
  pages = {406--461},
  doi = {10.1002/rsa.21063}
}

@article{abbe2018_CommunityDetectionStochastic,
  title = {Community {{Detection}} and {{Stochastic Block Models}}: {{Recent Developments}}},
  author = {Abbe, Emmanuel},
  year = 2018,
  journal = {Journal of Machine Learning Research},
  volume = {18},
  number = {177},
  pages = {1--86},
  url = {http://jmlr.org/papers/v18/16-480.html}
}

@article{newman2002_SpreadEpidemicDisease,
  title = {Spread of Epidemic Disease on Networks},
  author = {Newman, M. E. J.},
  year = 2002,
  month = jul,
  journal = {Physical Review E},
  volume = {66},
  number = {1},
  pages = {016128},
  doi = {10.1103/PhysRevE.66.016128}
}

@article{newman2009_RandomGraphsClustering,
  title = {Random {{Graphs}} with {{Clustering}}},
  author = {Newman, M. E. J.},
  year = 2009,
  month = jul,
  journal = {Physical Review Letters},
  volume = {103},
  number = {5},
  pages = {058701},
  publisher = {American Physical Society},
  doi = {10.1103/PhysRevLett.103.058701}
}

@article{gleeson2009_AnalyticalResultsBond,
  title = {Analytical Results for Bond Percolation and $k$-Core Sizes on Clustered Networks},
  author = {Gleeson, James P. and Melnik, Sergey},
  year = 2009,
  month = oct,
  journal = {Physical Review E},
  volume = {80},
  number = {4},
  pages = {046121},
  publisher = {American Physical Society},
  doi = {10.1103/PhysRevE.80.046121}
}

@article{penrose2016_ConnectivitySoftRandom,
  title = {Connectivity of Soft Random Geometric Graphs},
  author = {Penrose, Mathew D.},
  year = 2016,
  month = apr,
  journal = {The Annals of Applied Probability},
  volume = {26},
  number = {2},
  pages = {986--1028},
  publisher = {Institute of Mathematical Statistics},
  doi = {10.1214/15-AAP1110}
}

@article{mann2025_AlternativeExpressionMessage,
  title = {Alternative Expression of Message Passing on Networks},
  author = {Mann, Peter and Dobson, Simon},
  year = 2025,
  month = jun,
  journal = {Physical Review E},
  volume = {111},
  number = {6},
  pages = {064301},
  publisher = {American Physical Society},
  doi = {10.1103/PhysRevE.111.064301}
}

@misc{hutchcroft2025_CriticalLongrangePercolation1,
  title = {Critical Long-Range Percolation {{I}}: {{High}} Effective Dimension},
  author = {Hutchcroft, Tom},
  year = 2025,
  month = aug,
  eprint = {2508.18807},
  primaryclass = {math.PR},
  publisher = {arXiv},
  doi = {10.48550/arXiv.2508.18807},
  archiveprefix = {arXiv}
}

@misc{hutchcroft2025_CriticalLongrangePercolation2,
  title = {Critical Long-Range Percolation {{II}}: {{Low}} Effective Dimension},
  author = {Hutchcroft, Tom},
  year = 2025,
  month = aug,
  eprint = {2508.18808},
  primaryclass = {math.PR},
  publisher = {arXiv},
  doi = {10.48550/arXiv.2508.18808},
  archiveprefix = {arXiv}
}

@misc{hutchcroft2025_CriticalLongrangePercolation3,
  title = {Critical Long-Range Percolation {{III}}: {{The}} Upper Critical Dimension},
  author = {Hutchcroft, Tom},
  year = 2025,
  month = aug,
  eprint = {2508.18809},
  primaryclass = {math.PR},
  publisher = {arXiv},
  doi = {10.48550/arXiv.2508.18809},
  archiveprefix = {arXiv}
}

@article{peixoto2012_EvolutionRobustNetwork,
  title = {Evolution of {{Robust Network Topologies}}: {{Emergence}} of {{Central Backbones}}},
  author = {Peixoto, Tiago P. and Bornholdt, Stefan},
  year = 2012,
  month = sep,
  journal = {Physical Review Letters},
  volume = {109},
  number = {11},
  pages = {118703},
  publisher = {American Physical Society},
  doi = {10.1103/PhysRevLett.109.118703}
}

@article{sonmez2021_GraphDistancesContinuum,
  title = {Graph Distances of Continuum Long-Range Percolation},
  author = {S{\"o}nmez, Ercan},
  year = 2021,
  journal = {Brazilian Journal of Probability and Statistics},
  volume = {35},
  number = {3},
  pages = {609--624},
  doi = {10.1214/21-BJPS500}
}

@phdthesis{morelbalbi2022_LargeScaleStructure,
  title = {Large-Scale Structure of Multi-Optimised Networks},
  author = {Morel-Balbi, Sebastian},
  year = 2022,
  month = may,
  url = {https://purehost.bath.ac.uk/ws/portalfiles/portal/252092704/thesis.pdf},
  school = {University of Bath}
}

@article{newman2023_MessagePassingMethods,
  title = {Message Passing Methods on Complex Networks},
  author = {Newman, M. E. J.},
  year = 2023,
  month = feb,
  journal = {Proceedings of the Royal Society A: Mathematical, Physical and Engineering Sciences},
  volume = {479},
  number = {2270},
  pages = {20220774},
  doi = {10.1098/rspa.2022.0774},
  url = {https://doi.org/10.1098/rspa.2022.0774}
}

@article{newman2003_MixingPatternsNetworks,
  title = {Mixing Patterns in Networks},
  author = {Newman, M. E. J.},
  year = 2003,
  month = feb,
  journal = {Physical Review E},
  volume = {67},
  number = {2},
  pages = {026126},
  publisher = {American Physical Society},
  doi = {10.1103/PhysRevE.67.026126},
  url = {https://link.aps.org/doi/10.1103/PhysRevE.67.026126}
}

@article{olhede2014_NetworkHistogramsUniversality,
  title = {Network Histograms and Universality of Blockmodel Approximation},
  author = {Olhede, Sofia C. and Wolfe, Patrick J.},
  year = 2014,
  month = oct,
  journal = {Proceedings of the National Academy of Sciences},
  volume = {111},
  number = {41},
  pages = {14722--14727},
  publisher = {Proceedings of the National Academy of Sciences},
  doi = {10.1073/pnas.1400374111},
  url = {https://www.pnas.org/doi/abs/10.1073/pnas.1400374111}
}

@article{morel-balbi2020_NullModelsMultioptimized,
  title = {Null Models for Multioptimized Large-Scale Network Structures},
  author = {{Morel-Balbi}, Sebastian and Peixoto, Tiago P.},
  year = 2020,
  month = sep,
  journal = {Physical Review E},
  volume = {102},
  number = {3},
  pages = {032306},
  publisher = {American Physical Society},
  doi = {10.1103/PhysRevE.102.032306},
  url = {https://link.aps.org/doi/10.1103/PhysRevE.102.032306}
}

@misc{chebunin2026_StrongSharpPhase,
  title = {On Strong Sharp Phase Transition in the Random Connection Model},
  author = {Chebunin, Mikhail and Last, G{\"u}nter},
  year = 2025,
  eprint = {2512.00213},
  primaryclass = {math.PR},
  publisher = {arXiv},
  doi = {10.48550/arXiv.2512.00213},
  url = {http://arxiv.org/abs/2512.00213},
  archiveprefix = {arXiv}
}

@article{gori2017_OnedimensionalLongrangePercolation,
  title = {One-Dimensional Long-Range Percolation: {{A}} Numerical Study},
  author = {Gori, G. and Michelangeli, M. and Defenu, N. and Trombettoni, A.},
  year = 2017,
  month = jul,
  journal = {Physical Review E},
  volume = {96},
  number = {1},
  pages = {012108},
  publisher = {American Physical Society},
  doi = {10.1103/PhysRevE.96.012108},
  url = {https://link.aps.org/doi/10.1103/PhysRevE.96.012108}
}

@article{artime2024_RobustnessResilienceComplex,
  title = {Robustness and Resilience of Complex Networks},
  author = {Artime, Oriol and Grassia, Marco and De Domenico, Manlio and Gleeson, James P. and Makse, Hern{\'a}n A. and Mangioni, Giuseppe and Perc, Matja{\v z} and Radicchi, Filippo},
  year = 2024,
  month = feb,
  journal = {Nature Reviews Physics},
  volume = {6},
  number = {2},
  pages = {114--131},
  publisher = {Nature Publishing Group},
  doi = {10.1038/s42254-023-00676-y},
  url = {https://www.nature.com/articles/s42254-023-00676-y}
}

@article{jiang2025_RobustnessSmallNetworks,
  title = {Robustness of small Networks},
  author = {Jiang, Jessica and Zhuang, Allison C. and Holme, Petter and Mucha, Peter J. and Schwarze, Alice C.},
  year = 2026,
  month = aug,
  journal = {Physical Review E},
  volume = {114},
  number = {2},
  pages = {024307},
  publisher = {American Physical Society},
  doi = {10.1103/xq8k-864y},
  url = {https://doi.org/10.1103/xq8k-864y}
}

@inproceedings{bhamidi2026_StochasticBlockModel,
  title = {The {{Stochastic Block Model Has}} the {{Overlap Graph Property}} for {{Modularity}}},
  author = {Bhamidi, Shankar and Gamarnik, David and van der Hofstad, Remco and Litvak, Nelly and Pra{\l}at, Pawe{\l} and Skerman, Fiona and Tousinejad, Yasmin},
  year = 2026,
  booktitle = {53rd International Colloquium on Automata, Languages, and Programming (ICALP 2026)},
  series = {Leibniz International Proceedings in Informatics (LIPIcs)},
  volume = {374},
  pages = {28:1--28:20},
  publisher = {Schloss Dagstuhl -- Leibniz-Zentrum f{\"u}r Informatik},
  address = {Dagstuhl, Germany},
  doi = {10.4230/LIPIcs.ICALP.2026.28},
  url = {https://doi.org/10.4230/LIPIcs.ICALP.2026.28}
}

@article{schawe2020_LargeDeviationsConnected,
  title = {Large Deviations of Connected Components in the Stochastic Block Model},
  author = {Schawe, Hendrik and Hartmann, Alexander K.},
  year = 2020,
  month = nov,
  journal = {Physical Review E},
  volume = {102},
  number = {5},
  pages = {052108},
  publisher = {American Physical Society},
  doi = {10.1103/PhysRevE.102.052108},
  url = {https://link.aps.org/doi/10.1103/PhysRevE.102.052108}
}

@article{hoff2002_LatentSpaceApproaches,
  title = {Latent {{Space Approaches}} to {{Social Network Analysis}}},
  author = {Hoff, Peter D and Raftery, Adrian E and Handcock, Mark S},
  year = 2002,
  month = dec,
  journal = {Journal of the American Statistical Association},
  volume = {97},
  number = {460},
  pages = {1090--1098},
  publisher = {Taylor \& Francis},
  doi = {10.1198/016214502388618906},
  url = {https://doi.org/10.1198/016214502388618906}
}

@article{penrose1991_continuumpercolationmodel,
  title = {On a Continuum Percolation Model},
  author = {Penrose, Mathew D.},
  year = 1991,
  month = sep,
  journal = {Advances in Applied Probability},
  volume = {23},
  number = {3},
  pages = {536--556},
  doi = {10.2307/1427621},
  url = {https://www.cambridge.org/core/journals/advances-in-applied-probability/article/on-a-continuum-percolation-model/CA657E4DF36D3E6A45354A7A08F21149}
}

@article{penrose2022_Giantcomponentsoft,
  title = {Giant Component of the Soft Random Geometric Graph},
  author = {Penrose, Mathew D.},
  year = 2022,
  month = jan,
  journal = {Electronic Communications in Probability},
  volume = {27},
  eid = {53},
  pages = {1--10},
  publisher = {{Institute of Mathematical Statistics and Bernoulli Society}},
  doi = {10.1214/22-ECP491},
  url = {https://projecteuclid.org/journals/electronic-communications-in-probability/volume-27/issue-none/Giant-component-of-the-soft-random-geometric-graph/10.1214/22-ECP491.full}
}

@misc{kupper2026_Largestcomponentsharpness,
  title = {Largest Component and Sharpness in Continuum Percolation},
  author = {K{\"u}pper, Niclas and Penrose, Mathew D.},
  year = 2024,
  month = may,
  eprint = {2407.10715},
  primaryclass = {math.PR},
  publisher = {arXiv},
  doi = {10.48550/arXiv.2407.10715},
  url = {http://arxiv.org/abs/2407.10715},
  archiveprefix = {arXiv}
}

@misc{higgs2025_Exponentialdecayrandom,
  title = {Exponential Decay for the Random Connection Model Using Asymptotic Transitivity},
  author = {Higgs, Frankie},
  year = 2025,
  month = sep,
  eprint = {2509.02310},
  primaryclass = {math.PR},
  publisher = {arXiv},
  doi = {10.48550/arXiv.2509.02310},
  url = {http://arxiv.org/abs/2509.02310},
  archiveprefix = {arXiv}
}

@misc{caicedo2025_Sharpnesspercolationphase,
  title = {Sharpness of the Percolation Phase Transition for Weighted Random Connection Models},
  author = {Caicedo, Alejandro and Kolesnikov, Leonid},
  year = 2025,
  month = dec,
  number = {arXiv:2512.21742},
  eprint = {2512.21742},
  primaryclass = {math.PR},
  publisher = {arXiv},
  doi = {10.48550/arXiv.2512.21742},
  url = {http://arxiv.org/abs/2512.21742},
  archiveprefix = {arXiv}
}

@article{hutchcroft2021_Powerlawboundscritical,
  title = {Power-Law Bounds for Critical Long-Range Percolation below the Upper-Critical Dimension},
  author = {Hutchcroft, Tom},
  year = 2021,
  month = nov,
  journal = {Probability Theory and Related Fields},
  volume = {181},
  number = {1},
  pages = {533--570},
  doi = {10.1007/s00440-021-01043-7},
  url = {https://doi.org/10.1007/s00440-021-01043-7}
}

@article{schulman1983_Longrangepercolation,
  title = {Long Range Percolation in One Dimension},
  author = {Schulman, L. S.},
  year = 1983,
  month = dec,
  journal = {Journal of Physics A: Mathematical and General},
  volume = {16},
  number = {17},
  pages = {L639},
  doi = {10.1088/0305-4470/16/17/001},
  url = {https://doi.org/10.1088/0305-4470/16/17/001}
}

@article{newman1986_Onedimensional1,
  title = {One Dimensional $1/|j - i|^s$ Percolation Models: The Existence of a Transition for $s \le 2$},
  author = {Newman, C. M. and Schulman, L. S.},
  year = 1986,
  month = dec,
  journal = {Communications in Mathematical Physics},
  volume = {104},
  number = {4},
  pages = {547--571},
  doi = {10.1007/BF01211064},
  url = {https://doi.org/10.1007/BF01211064}
}

@article{aizenman1986_Discontinuitypercolationdensity,
  title = {Discontinuity of the Percolation Density in One Dimensional $1/|x-y|^2$ Percolation Models},
  author = {Aizenman, M. and Newman, C. M.},
  year = 1986,
  month = dec,
  journal = {Communications in Mathematical Physics},
  volume = {107},
  number = {4},
  pages = {611--647},
  doi = {10.1007/BF01205489},
  url = {https://doi.org/10.1007/BF01205489}
}

@misc{coupette2025_universallyapplicableapproach,
  title = {A Universally Applicable Approach to Connectivity Percolation},
  author = {Coupette, Fabian and Schilling, Tanja},
  year = 2023,
  month = jun,
  eprint = {2308.16757},
  primaryclass = {cond-mat.stat-mech},
  publisher = {arXiv},
  doi = {10.48550/arXiv.2308.16757},
  url = {http://arxiv.org/abs/2308.16757},
  archiveprefix = {arXiv},
}

@article{miller2009_Percolationepidemicsrandom,
  title = {Percolation and Epidemics in Random Clustered Networks},
  author = {Miller, Joel C.},
  year = 2009,
  month = aug,
  journal = {Physical Review E},
  volume = {80},
  number = {2},
  pages = {020901},
  publisher = {American Physical Society},
  doi = {10.1103/PhysRevE.80.020901},
  url = {https://link.aps.org/doi/10.1103/PhysRevE.80.020901}
}

@article{callaway2000_NetworkRobustnessFragility,
  title = {Network {{Robustness}} and {{Fragility}}: {{Percolation}} on {{Random Graphs}}},
  author = {Callaway, Duncan S. and Newman, M. E. J. and Strogatz, Steven H. and Watts, Duncan J.},
  year = 2000,
  month = dec,
  journal = {Physical Review Letters},
  volume = {85},
  number = {25},
  pages = {5468--5471},
  publisher = {American Physical Society},
  doi = {10.1103/PhysRevLett.85.5468},
  url = {https://link.aps.org/doi/10.1103/PhysRevLett.85.5468}
}

@article{molloy1995_criticalpointrandom,
  title = {A Critical Point for Random Graphs with a given Degree Sequence},
  author = {Molloy, Michael and Reed, Bruce},
  year = 1995,
  journal = {Random Structures \& Algorithms},
  volume = {6},
  number = {2-3},
  pages = {161--180},
  doi = {10.1002/rsa.3240060204},
  url = {https://onlinelibrary.wiley.com/doi/abs/10.1002/rsa.3240060204}
}

@article{spencer2023_Projectivesparselearnable,
  title = {Projective, Sparse and Learnable Latent Position Network Models},
  author = {Spencer, Neil A. and Shalizi, Cosma Rohilla},
  year = 2023,
  month = dec,
  journal = {The Annals of Statistics},
  volume = {51},
  number = {6},
  pages = {2506--2525},
  publisher = {Institute of Mathematical Statistics},
  doi = {10.1214/23-AOS2340},
  url = {https://projecteuclid.org/journals/annals-of-statistics/volume-51/issue-6/Projective-sparse-and-learnable-latent-position-network-models/10.1214/23-AOS2340.full}
}

\end{document}